\documentclass[a4paper,11pt]{article}
\pdfoutput=1 

\usepackage{jcappub} 

\usepackage[T1]{fontenc} 
\usepackage{natbib}
\usepackage{amsmath}
\usepackage{bbold}

\usepackage[colorinlistoftodos]{todonotes}

\title{\boldmath Forecasting Super-Sample Covariance in Future Weak Lensing Surveys with SuperSCRAM}

\author{Matthew C. Digman,}
\author{Joseph E. McEwen, and}
\author{Christopher M. Hirata}

\affiliation{Center for Cosmology and AstroParticle Physics, Department of Physics, The Ohio State University, 191 W Woodruff Ave, Columbus OH 43210, USA}

\emailAdd{digman.12@osu.edu}

\abstract{The observable universe contains density perturbations on scales larger than any finite volume survey. Perturbations on scales larger than a survey can measure degrade its power to constrain cosmological parameters. The dependence of survey observables such as the weak lensing power spectrum on these long-wavelength modes results in super-sample covariance. Accurately forecasting parameter constraints for future surveys requires accurately accounting for the super-sample effects. If super-sample covariance is in fact a major component of the survey error budget, it may be necessary to investigate mitigation strategies that constrain the specific realization of the long-wavelength modes. We present a Fisher matrix based formalism for approximating the magnitude of super-sample covariance and the effectiveness of mitigation strategies for realistic survey geometries. We implement our formalism in the public code SuperSCRAM: Super-Sample Covariance Reduction and Mitigation. We illustrate SuperSCRAM with an example application, where the modes contributing to super-sample covariance in the WFIRST weak lensing survey are constrained by the low-redshift galaxy number counts in the wider LSST footprint. We find that super-sample covariance increases the volume of the error ellipsoid in 7D cosmological parameter space by a factor of 4.5 relative to Gaussian statistical errors only, but our simple mitigation strategy more than halves the contamination, to a factor of 2.0.}


\newcommand{\sph}[2]{Y^\text{R}_{l_#1 m_#1}(\hat{#2})}

\newcommand{\jl}[1]{j_{l_#1}}
\newcommand{\dk}{\frac{ d^3 \mathbf{k}}{(2 \pi)^3}} 
\newcommand{ \dkv}[1]{\frac{ d^3 \mathbf{k}_{#1}}{(2 \pi)^3}}

\begin{document}
\maketitle
\flushbottom

\sloppy 
\section{Introduction}
\label{sec:intro}

Observational cosmology has experienced a remarkable rebirth as a precision science in the last two decades. In 1998, two teams using Type Ia supernovae found that the expansion of the Universe is accelerating \cite{Perlmutter:1998np,Riess:1998cb}.
Galaxy surveys such as the 2dF Galaxy Redshift Survey \cite{2df2005}, SDSS \cite{sdss}, BOSS \cite{boss2013}, and WiggleZ \cite{wigglez2012} have provided an independent means of exploring the recent expansion history of the Universe and have dramatically confirmed cosmic acceleration.
Meanwhile, the high-redshift Universe has been probed to great precision using the full sky surveys of the cosmic microwave background by WMAP \cite{wmap1year,wmap9year} Planck \cite{planck2015params,planck_2018}, ACT \cite{act_3year}, and SPT \cite{spt_cosmo}.

The cause of the accelerated expansion of the universe is tied to the origin and ultimate fate of the universe. Depending on how the acceleration changes over time, the universe could expand forever, re-collapse, rip itself apart, or perhaps do something else entirely \cite{bigripcaldwell,fate_universe}. The most common parametrization for the cause of cosmic acceleration introduces a fluid with positive energy density but negative pressure, called dark energy.  Dark energy models are characterized by an equation of state parameter $w\equiv \frac{p}{\rho}$, which relates its pressure $p$ and the energy density $\rho$ \cite{constraining_de}. In the simple case where $w=-1$, dark energy is a cosmological constant, which is the standard $\Lambda$CDM cosmological model. The most studied physical mechanisms which could produce a $w\neq-1$ would also allow $w$ to vary as a function of time, $w=w(z)$ \cite{quintessence_caldwell}. Current cosmological observations do not give good constraints on $w(z)$ because errors on $w(z)$ at different times are highly correlated \cite{weinberg_probes,eos_pitfalls}. 

Tomographic weak lensing surveys directly probe the statistical properties of the matter distribution in projection, using sources at a range of redshifts, and hence can be used to constrain $w(z)$ \cite{weinberg_probes,huterer_wl}. Several completed and ongoing experiments have measured or will measure weak lensing, including CFHTLens \cite{cfhtlens}, KiDS \cite{kids}, DES \cite{des}, and HSC \cite{hsc-wl}.
Ideally, the precision of weak lensing measurements would be limited primarily by fundamental statistical uncertainties due to their limited samples of galaxies and survey volumes. Such statistical uncertainties will be greatly reduced by the large volumes of future surveys, such as LSST \cite{lsst}, Euclid \cite{euclid}, and WFIRST \cite{wfirst}. However, future weak lensing surveys will also suffer from many sources of observational and astrophysical systematic errors, which will need to be properly understood to avoid systematic biases in the results \cite{schaan_shear_calibration,systematic_lensing_mandelbaum,systematic_lensing_massey,systematic_lensing_huterer}. One source of statistical error for future experiments to consider is the coupling of measured results to matter density fluctuations outside the survey window, called super-sample covariance (SSC). The direct coupling between the amplitude of short-wavelength modes, which fit in the survey window, and long-wavelength modes, which do not, was described as beat-coupling in refs.~\cite{hamilton_ssc_orig1,hamilton_ssc_orig2} and found to be the dominant contribution to the matter power spectrum covariance on small scales. Further studies have considered other aspects of the effect and described formalism for analyzing it \cite{lihu_ssc_signal,lihu_ssc_sim,cov_jackknife,takadahu_ssc,takadaspergel_ssc,deputter_ssc}. The SSC effect on lensing observations has been shown to depend strongly on the details of the survey geometry \cite{geom_wl_ssc}. Survey geometry design is subject to many practical instrument constraints, such as observatory slew times, galactic plane avoidance, and calibration. Future experiments, such as WFIRST, should consider whether optimizing survey design to reduce super-sample covariance is a significant additional consideration.  Additionally, missions should consider whether combining other probes of long-wavelength density fluctuations can be a useful means of mitigating their covariances \cite{krause_eifler_cosmolike,eifler_krause_cosmolike}. 

In this paper, we present a formalism for estimating the SSC contribution to the covariances of future experiments and investigating possible mitigation strategies in a realistic survey geometry. We implement our formalism in a public code, SuperSCRAM: Super-Sample Covariance Reduction and Mitigation.\footnote{SuperSCRAM is available at \url{https://github.com/mcdigman/SuperSCRAM}} The code is intended to be modular and extensible to facilitate the addition of further observables, physical effects, and mitigation strategies. SuperSCRAM provides Fisher matrix forecasts of parameter constraints, which is useful for estimation of the relative magnitude of effects and effectiveness of mitigation strategies, but is not a substitute for a full Markov chain Monte Carlo likelihood code such as CosmoLike \cite{krause_eifler_cosmolike,eifler_krause_cosmolike}. We do not analyze other sources of non-Gaussian covariance, although SuperSCRAM could be extended to include them.

Refs.~\cite{lacasa_cluster_counts,lacasa_partial_sky,lacasa_ssc,lacasa_angular} have applied a similar harmonic expansion formalism to calculate SSC effects on cluster counts including geometrical effects. This paper focuses on weak lensing observables, although the code's structure is modular, so an extension to include cluster counts in the future would be straightforward.

The structure of the paper is as follows. In section~\ref{sec:Overview}, we describe the formalism and overall analysis workflow used throughout the paper. In section~\ref{sec:method}, we describe the specific implementation of our formalism in SuperSCRAM. In subsection~\ref{ssec:example}, we present a practical demonstration of a useful application of SuperSCRAM, in which we assess the effectiveness of embedding the WFIRST weak lensing survey program in the LSST field of view. In subsection~\ref{ssec:survey_geometry} and subsection~\ref{ssec:extensions}, we discuss other applications and extensions of SuperSCRAM. Finally, in section~\ref{sec:conclusion} we discuss our conclusions and the outlook for future work. 

\section{Overview}
\label{sec:Overview}

\subsection{Notation and Definitions}

For our analysis, we define two types of observables, which must be handled in separate ways: short-wavelength (SW) observables, denoted $\{O^I\}$, and long-wavelength (LW) observables, denoted $\{O^a\}$. The cosmological parameters are denoted $\{\Theta^i\}$. Finally, the long-wavelength matter density modes are expanded in terms of a set of basis modes $\{\psi_\alpha({\bf r})\}$, $\delta({\bf r}) = \sum_\alpha \delta^\alpha \psi_\alpha({\bf r})$. In this paper, we use indices $IJK...$ for SW observables, $abc...$ for LW observables, $ijk...$ for cosmological parameters, and $\alpha\beta\gamma...$ for the density mode basis.

Short-wavelength observables, discussed in subsection~\ref{ssec:sw_observables}, are observables which can only directly probe fluctuations of shorter wavelength than the survey's window, such as the shear-shear lensing power spectrum, or cluster counts. In this paper, they are the ``primary'' source of cosmological information, which a survey is designed to extract.

Long-wavelength matter density fluctuations, with wavelengths larger than the survey window, add covariance to estimators of the short-wavelength modes measured within the window, which ultimately degrades cosmological parameter estimation. Long-wavelength observables, discussed in subsection~\ref{ssec:lw_observables}, are chosen to provide information about the amplitude of the long-wavelength matter density fluctuations, so the degradation of estimates of short-wavelength observables by super-sample covariance can be mitigated. Long-wavelength observables need not originate from the same survey that measures the short-wavelength observables.

\subsection{Workflow}

\begin{figure}[!ht]
 \centering
\includegraphics[width=1\textwidth]{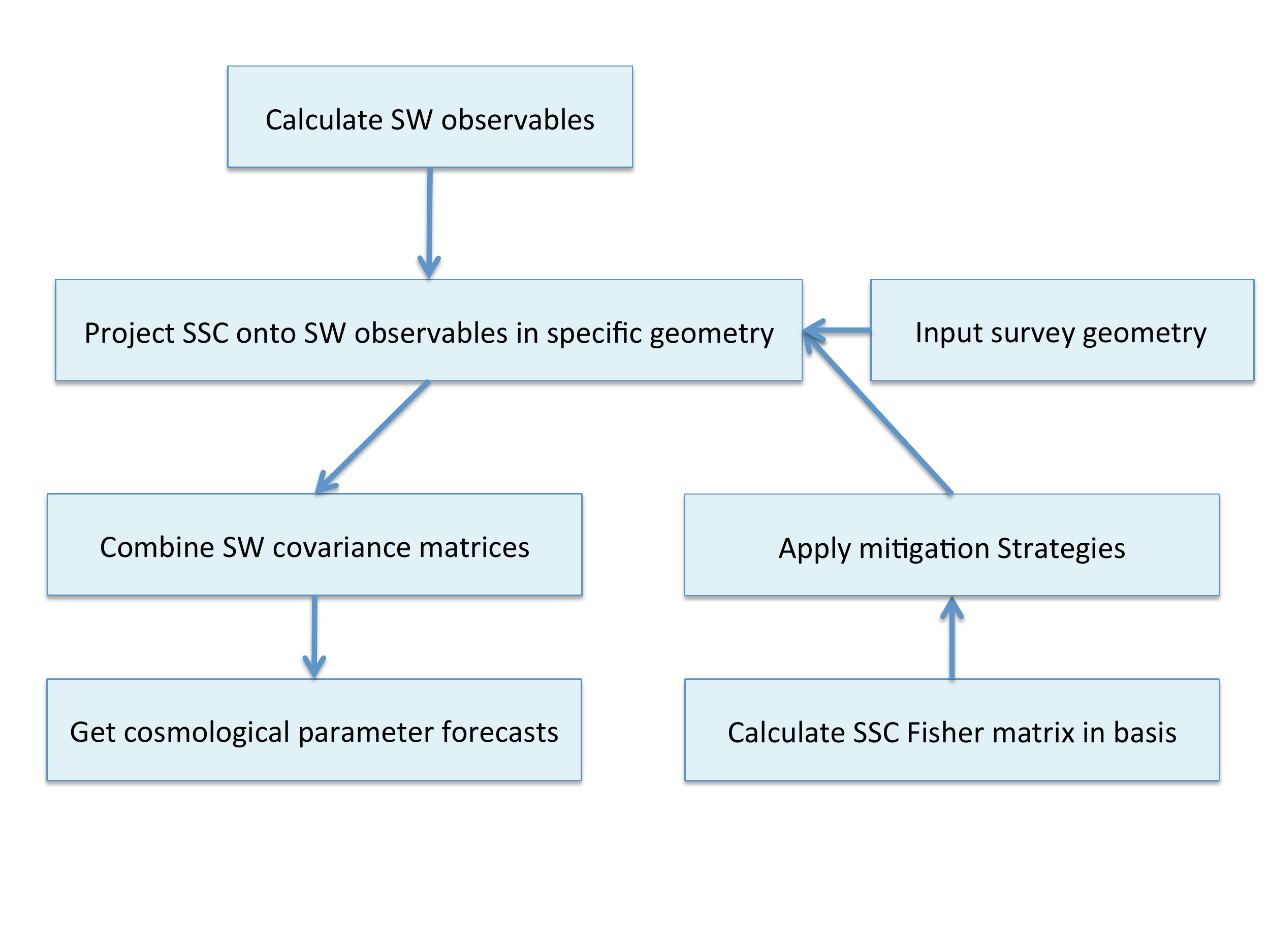}
  \caption{Conceptual overview of the basic workflow needed to obtain cosmological parameter forecasts. The specific modules implementing the structure are described in section~\ref{sec:method} and shown in figure~\ref{fig:codeflow}.}
\label{fig:workflow}
\end{figure}

A sketch of the workflow of our mitigation procedure is provided in figure~\ref{fig:workflow}.

To provide cosmological parameter forecasts, we calculate the total Fisher information matrix for the cosmological parameters by combining prior constraints and the information obtained from the short-wavelength observables measured by the survey:
\begin{align}\label{fisher_param}
F_{ij}^{\text{tot}}=F_{ij}^{\text{prior}}+F_{ij}^{\text{survey}},
\end{align}
where the indices $i$ and $j$ run over all possible combinations of cosmological parameters. For the purposes of this paper, we use the WMAP5 fiducial point with a set of forecast constraints from Planck for $F_{ij}^{\text{prior}}$, taken from ref.~\cite{jdem_fom}. To calculate  $F_{ij}^{\text{survey}}$, we use
\begin{align}\label{fisher_survey}
F_{ij}^{\text{survey}}=\frac{\partial O^I}{\partial \Theta^i} F_{IJ}^{\text{tot}}\frac{\partial O^J}{\partial \Theta^j},
\end{align}
 where here $O^I$ are the short-wavelength observable vectors. $F_{IJ}^{\text{tot}}=\left(C^{IJ}_{\text{tot}}\right)^{-1}$ is calculated as the inverse of the short-wavelength covariance matrix, 
\begin{align}\label{cov_sw_tot}
C^{IJ}_{\text{tot}}=C^{IJ}_{\text{g}}+C^{IJ}_{\text{ng}}+C^{IJ}_{\text{SSC}}.
\end{align}
In this paper, we do not consider the non-Gaussian covariance $C^{IJ}_{\text{ng}}$, as our analysis is primarily focused on estimating the contribution of the super-sample covariance $C^{IJ}_{\text{SSC}}$ term. We decompose the long-wavelength modes into a basis as described in Appendix~{\ref{app:basis}}, and calculate the super-sample covariance in our basis using perturbation theory, such that we can write
\begin{align}\label{cov_ssc}
C^{IJ}_{\text{SSC}} = \frac{\partial O^I}{\partial \delta^\alpha} (F^{\alpha\beta}_{\text{SSC}})^{-1}\frac{\partial O^J}{\partial \delta^\beta}. 
\end{align}

 The response of the observable to a density fluctuation $\partial O^I/\partial \delta^\alpha$ is described in subsection~\ref{ssec:obs_response}. $F_{\alpha\beta}^\text{SSC}$ is the Fisher information matrix for long-wavelength matter density fluctuations decomposed into our basis, which includes the SSC contribution from perturbation theory given in eq.~\eqref{C_SSC}, plus any information provided by mitigation strategies, 
\begin{align}\label{F_tot}
F_{\alpha\beta}^{\text{SSC+mit}} = F_{\alpha\beta}^{\text{SSC}}+\sum_{n}{F_{\alpha\beta}^{n}}.
\end{align}
The response to the long-wavelength fluctuations and the Fisher information matrices of long-wavelength observables contain explicit dependence on the geometry of the survey. Note that eq.~\eqref{F_tot} implicitly assumes that the long-wavelength observables are statistically independent on large scales. Statistical independence is generally a reasonable assumption if the two observables are coming from different types of measurements or non-overlapping surveys. However, care must be taken to avoid adding the same information twice if observables contain information from overlapping datasets. 

It is worth noting that the SSC term is not purely an observational limitation from partial sky coverage, and $C^{IJ}_{\text{SSC}}$ would not vanish even for a full sky survey, because an important component of the SSC term is radial. In fact, the improvements to the Gaussian covariance from increasing the sky area may make the SSC term relatively more important. Therefore, the SSC term is an important component of the error budget even for large surveys \cite{2018JCAP...10..053B}. The specific mitigation approach implemented in this paper --- embedding a survey in a larger survey --- would not work in this case, but it is possible that other approaches would.

\subsection{Eigenvalue analysis}
\label{ssec:eig}

Inspecting the individual values of the elements of the covariance matrix of the cosmological parameters $C^{ij}_{\text{tot}}$ can give a rough idea of the magnitude of the SSC contamination or effectiveness of a mitigation strategy. However, a more systematic method is required to quantitatively assess the overall impact. The direction of parameter space most contaminated by SSC or improved by mitigation is very difficult to determine by visual inspection of two-dimensional slices of many-dimensional ellipsoids, such as figure~\ref{fig:ellipse_plot}. 

To assess the overall impact, we use one covariance matrix, as a metric to identify the directions most changed in another perturbed covariance matrix. For our purposes,  a useful metric is the purely Gaussian covariance in cosmological parameter space, $C^{ij}_\text{g}$. For this section, we will refer to the covariance in cosmological parameter space with the super-sample contribution but no mitigation as $C^{ij}_\text{SSC}$, and  the covariance with all mitigation applied as $C^{ij}_\text{mit}$. We then consider the product matrix
\begin{equation}
\label{m_tot_gauss}
\mathcal{M}^i_{\;j}=(C^{ik}_{\text{SSC}})(C_\text{g}^{-1})_{kj},
\end{equation}
which has eigenvalues $\lambda^{(i)}$ and eigenvectors $v^i_{\;j}$ such that $\mathcal{M}^i_{\;j} v^{j(k)}=\lambda^{(k)} v^{i(k)}$. For  this section, Latin indices $ijk...$ run over cosmological parameter space and must be raised or lowered by applying the $C_\text{g}^{ij}$ or its inverse $(C_\text{g}^{-1})_{ij}$. Indices in parentheses $(i)(j)(k)...$ are labels for the eigenvalues and eigenvectors, and are not summed over here.  If $C^{ij}_{\text{SSC}}$ is a positive semidefinite perturbation to the metric, all of the eigenvalues satisfy $\lambda^{(i)}\ge1$, and the largest eigenvalue corresponds to the direction in parameter space most contaminated by $C_{\text{SSC}}^{ij}$. The resulting eigenvalues are dimensionless, and can therefore be compared between cosmological parametrizations. Note that the matrix $\mathcal{M}^i_{\;j}$ is not symmetric in general, which is inconvenient for computing eigenvalues numerically, so we instead find the eigenvalues of the matrix
\begin{equation}\label{m_tot_alt}
\widetilde{\mathcal{M}}_{\hat{i}\hat{j}} = (L^{-1})_{k\hat{i}}C_{\text{SSC}}^{kl}(L^{-1})_{l\hat{j}},
\end{equation}
where $L^{i\hat{j}}$ is the lower triangular Cholesky decomposition $C_\text{g}^{ij}\equiv L^{i\hat{k}}L^{j\hat{k}}$. $\widetilde{\mathcal{M}}_{\hat{i}\hat{j}}$ is a symmetric matrix with the same eigenvalues $\lambda^{(i)}$ as $\mathcal{M}^i_j$ and eigenvectors $u_{\hat{i}}^{\;(j)}$ related to the eigenvectors $v^{i(j)}$ of $\mathcal{M}^i_{\;j}$ by transforming $u_{\hat{i}}^{\;(j)}=(L^{-1})_{k\hat{i}} v^{k(j)}$. The transformed eigenvectors then exist in a new vector space where the metric is $\mathbb{1}$. Indices in this vector space, $\hat{i}\hat{j}\hat{k}...$, may therefore be raised and lowered freely.

Now, we would like to interpret the most contaminated directions in our physical parameter space. We define the one-form $\tilde{v}_{i}^{\;(j)}\equiv (C_\text{g}^{-1})_{ik}v^{k(j)}$. $\tilde{v}_{i}^{\;(j)}$ has the property $\tilde{v}_{j}^{\;(i)} C_{\text{SSC}}^{jk}\tilde{v}_{k}^{\;(l)}=u^{\hat{j}(i)}\widetilde{\mathcal{M}}_{\hat{j}\hat{k}}u^{\hat{k}(l)}$, such that $\tilde{v}_{i}^{\;(j)}$ gives the directions in the physical parameter space most contaminated by super-sample covariance relative to the Gaussian covariance. We require the normalization $u^{\hat{k}(i)} u_{\hat{k}}^{(j)}\equiv\mathbb{1}^{(i)(j)}$.

The covariance matrix in the basis of directions most contaminated by super-sample covariance is given:
\begin{equation}\label{oneform_covariance}
\widetilde{C}^{(i)(j)}_{\text{SSC},\text{mit}} \equiv \tilde{v}^{(i)}_{\;k} C^{kl}_{\text{mit}}\tilde{v}^{(j)}_{l}, 
\end{equation}
where $C_{\text{mit}}^{ij}$ is the covariance matrix with both super-sample covariance and mitigation applied. The elements of  $\widetilde{C}^{(i)(j)}_{\text{SSC},\text{mit}}$ give the covariance relative to $C_\text{g}^{ij}$ in the directions most contaminated without mitigation. If mitigation is perfect, $C_\text{mit}^{ij}=C_\text{g}^{ij}$, we have $\widetilde{C}^{(i)(j)}_{\text{mit},C_\text{g}}=\mathbb{1}^{(i)(j)}$. If instead the mitigation had no effect,  $C^{ij}_\text{mit}=C^{ij}_\text{SSC}$, we have $\widetilde{C}^{(i)(j)}_{\text{SSC},\text{SSC}}=\lambda^{(i)}\mathbb{1}^{(i)(j)}$.

The product of the eigenvalues is the ratio of the volumes of the uncertainty ellipsoid with and without contamination, $\prod_{(i)}\lambda^{(i)}={\rm det}({\bf C}_\text{SSC})/{\rm det}({\bf C}_\text{g})$. Note that because changing the cosmological parametrization affects both $C^{ij}_\text{SSC}$ and $C_\text{g}^{ij}$ in the same way, $\prod_{(i)}\lambda^{(i)}$ and the largest eigenvalue  $\lambda^{\text{top}}$ are  both relatively insensitive to the choice of parametrization, which makes them useful for examining the effect of super-sample covariance in isolation. The product of eigenvalues and top eigenvalue are two useful points of comparison between different survey and mitigation strategies.

In subsection~\ref{ssec:example}, we examine the eigenvalues most contaminated by the addition of the SSC term before and after mitigation, $\lambda^{(i)}_{\text{SSC}}$, and $\lambda^{(i)}_{\text{mit}}$.

\section{Methods and Code Organization}
\label{sec:method}

\begin{figure}[!ht]
 \centering
\includegraphics[width=1\textwidth]{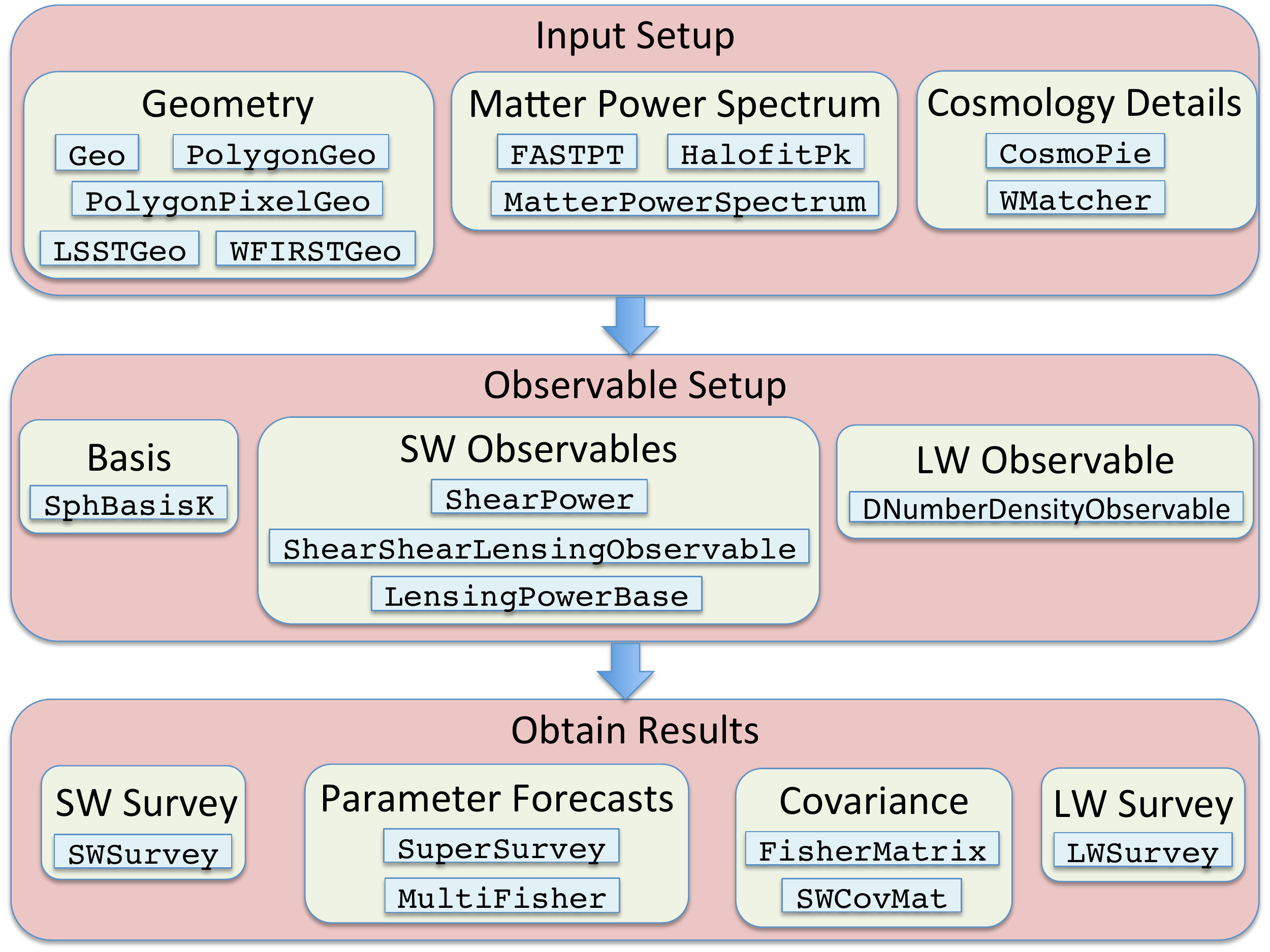}
  \caption{Structure of the most important modules used by SuperSCRAM described in this paper to implement the workflow described in \S\ref{fig:workflow}.}
\label{fig:codeflow}
\end{figure}

In this section, we present a framework and code for forecasting super-sample effects on future observations in in a survey with a given geometry. SuperSCRAM is written and tested in Python 2.7 and uses the SciPy \cite{scipy}, NumPy \cite{numpy}, and Astropy \cite{astropy_1,astropy_2} libraries. Convergence tests were run on the CCAPP condos of the Pitzer and Ruby clusters at the Ohio Supercomputer Center \cite{osc}. The overall structure of the important modules in SuperSCRAM is sketched out in figure~\ref{fig:codeflow}. 

\subsection{Geometries}
\label{ssec:geometries}

The base class for survey geometries is the \texttt{Geo} class, which contains basic functionality for setting up the tomographic lensing bins and the resolution bins to be integrated over. Classes that extend \texttt{Geo} for a specific angular window function must provide a method for calculating the angular area in steradians, \texttt{angular\_area()}, and the spherical harmonic coefficients $a_{l m}$, \texttt{a\_lm(l,m)}. 

We provide several possible methods of implementing survey geometries. Pixelated geometries extend the \texttt{PixelGeo} class, which allows for an arbitrary pixelation of the sky, though all current geometries use HEALPix \cite{HEALPix}. Our \texttt{PolygonGeo} class allows a survey geometry to be input as an arbitrary bounding spherical polygon, with segments of great circles as edges. The primary advantage of this class is that the spherical harmonic decomposition $a_{l m}$ can be calculated analytically for such polygons, as described in Appendix~\ref{app_analy_poly}. For comparison with pixelated methods also provide the \texttt{PolygonPixelGeo} class, which implements spherical polygons using a HEALPix pixelation.

We have implemented several utility geometries which perform useful operations on other geometries. The \texttt{AlmDifferenceGeo} implements the difference between any two \texttt{Geo} objects, which is necessary for cutting a mask out of a survey window. \texttt{AlmRotationGeo} uses the Euler rotation method described in Appendix~\ref{app_analy_poly} to rotate any \texttt{Geo} to a specified position on the sky. \texttt{PolygonUnionGeo} and \texttt{PolygonPixelUnionGeo} calculate the union of \texttt{PolygonGeo}  and \texttt{PolygonPixelGeo} objects respectively; we have not currently implemented a more general purpose class for calculating unions or intersections between two arbitrary geometries. 

We also provide a number of demonstration and testing geometries in \texttt{premade\_geos.py}. For our demonstration we use the \texttt{LSSTGeo} and \texttt{WFIRSTGeo}, which implement approximations of possible LSST and WFIRST window functions using \texttt{PolygonGeo} objects, combined using the various utility geometries. Various other geometries used for the demonstrations shown in figure~\ref{fig:sample_geos} include \texttt{StripeGeo}, which describes a spherical rectangle, and \texttt{CircleGeo}, which  approximates a circle as a many-sided spherical polygon. For testing purposes, we have implemented \texttt{FullSkyGeo} and \texttt{HalfSkyGeo}, which use the analytic $a_{l m}$ for the full sky or an entire hemisphere respectively.

Various utility functions, such as those needed to reconstruct the images of geometries in figure~\ref{fig:sample_geos} from their $a_{l m}$ representations, are provided in \texttt{alm\_utils.py}, \texttt{ylm\_utils.py}, and \texttt{geo\_display\_utils.py}.

\subsection{Short-Wavelength Observables}
\label{ssec:sw_observables}

The \texttt{SWObservable} class provides an interface which implementations of specific short-wavelength observables must provide. A short-wavelength observable must implement the method \texttt{get\_dO\_I\_ddelta\_bar}, which should calculate ${\partial O^I}/{\partial \bar{\delta}}$ and return it as a NumPy array. All short-wavelength observables currently implemented in SuperSCRAM are subclasses of  \texttt{LensingObservable}, which are projected power spectra calculated by \texttt{ShearPower}, as described in Appendix~\ref{app:projected_power}. We treat lensing between each pair of tomographic bins as separate \texttt{LensingObservable} objects. To avoid unnecessarily recomputing power spectra in different tomographic bins, each \texttt{SWSurvey} object creates a single shared \texttt{LensingPowerBase} object to store all the \texttt{ShearPower} objects which its \texttt{LensingObservable} objects will need.  

To create a specific lensing observable, a \texttt{LensingObservable} subclass, such as the shear-shear lensing power spectrum \texttt{ShearShearLensingObservable}, needs to be provided with two \texttt{QWeight} objects, in this case \texttt{QShear}, which are the functions $q_A(z)$ described in Appendix~\ref{app:projected_power}. 

\subsection{Long-Wavelength Observables}
\label{ssec:lw_observables}
In this section, we describe a sample mitigation strategy of embedding a weak lensing survey, such as WFIRST, in a wider but shallower survey of galaxy number counts, such as LSST. The difference between the number density measured by the larger survey inside the lensing survey window, $n_1$, and the number density outside, $n_2$, provides information about long-wavelength density fluctuations. Our observable is therefore $\Delta n_{12}\equiv n_1-n_2$.  The number density can be written to linear order in perturbation theory as
\begin{align}
n_1 &= \frac{3}{(r_\text{max}^3-r_\text{min}^3)\Omega_1}\int_{\Omega_1}d\Omega\int_{r_\text{min}}^{r_\text{max}}r^2 dr n(r)\left[1+b(r)\delta(r,\Omega)\right]
\nonumber\\
&= \frac{3}{(r_\text{max}^3-r_\text{min}^3)\Omega_1}\int_{\Omega_1}d\Omega\int_{r_\text{min}}^{r_\text{max}}r^2 dr n(r)\left[1+b(r)\sum_\alpha{\delta^\alpha\psi_\alpha(r,\Omega)}\right],
\label{number_density}
\end{align}
where $b(r)$ is the linear bias, $n(r)$ is the expected number density of galaxies at comoving distance $r$, $\Omega_1$ is the angular area of the window, and $\delta(r,\Omega)=\sum_\alpha{\delta^\alpha\psi_\alpha(r,\Omega)}$ is a density fluctuation expanded in our basis described in Appendix~\ref{app:basis}. 

For the number density of galaxies as a function of redshift in the mitigation survey, we use the expected number densities for the LSST 1-year survey to a depth of $i<24$ from ref.~\cite{lsst_sci_req}, which gives approximately $18$ galaxies/arcmin$^2$. We have deliberately chosen a very conservative use of LSST. We obtain the bias by abundance matching to the Sheth-Tormen halo mass function, as described in Appendix~\ref{app:nz_candels}.
Assuming the wider survey has uniform sensitivity as a function of sky position, we have the response to long-wavelength density perturbations:
\begin{align}
\frac{\partial \Delta n_{12}}{\partial \delta^\alpha}&\equiv \frac{\partial n_1}{\partial \delta^\alpha}-\frac{\partial n_2}{\partial \delta^\alpha} \label{number_density_response}
\\
&=  \frac{3}{r_\text{max}^3-r_\text{min}^3}\int_{r_\text{min}}^{r_\text{max}}r^2\, dr\, n(r)b(r)\left[\frac{1}{\Omega_1}\int_{\Omega_1}d\Omega \psi_{\alpha}(r,\Omega)-\frac{1}{\Omega_2}\int_{\Omega_2}d\Omega\psi_{\alpha}(r,\Omega)\right].\nonumber
\end{align}
The ``noise'' in the long-wavelength observable is the shot noise, since for our application the clustering of the galaxies due to the underlying large-scale matter density field is the signal of interest. The mean number density of galaxies in this slice is 
\begin{equation}
\left<n\right>=\frac{3}{(r_\text{max}^3-r_\text{min}^3)}\int_{r_\text{min}}^{r_\text{max}}n(r)r^2dr.
\end{equation}
We can calculate the variance under the assumption that the shot noise is Poissonian:
\begin{align}\label{number_density_variance}
N_{12}\equiv {\rm Var}\left(n_1-n_2\right)= \left<n\right>\left(\frac{1}{V_1}+\frac{1}{V_2}\right),
\end{align}
where $V_i \equiv (r_\text{max}^3-r_\text{min}^3)\Omega_i/3$. In real samples, the long-wavelength shot noise may deviate from the Poisson prediction due to non-linear clustering, exclusion, and satellite galaxies (see, e.g., ref.~\cite{2013PhRvD..88h3507B} for an extensive discussion). For our example application in section~\ref{sec:applications}, the mitigation performance is only weakly dependent on $N_{12}$.\footnote{Increasing the galaxy number count survey's limiting magnitude from $i<24.1$ to $i<25.1$ more than doubles the number of galaxies, but reduces the volume of the error ellipsoid in our 7D cosmological parameter space by $<1\%$.}

Then, we can write the Fisher matrix for this mitigation strategy
\begin{align}\label{lw_ssc}
F^\text{mit}_{\alpha\beta}=\frac{\partial \Delta n_{12}}{\partial \delta^\alpha}(N_{12})^{-1}\frac{\partial \Delta n_{12}}{\partial \delta^\beta}.
\end{align}

This long-wavelength observable is implemented in the \texttt{DNumberDensityObservable} class, which extends the general specifications for a long-wavelength observable, \texttt{LWObservable}. Subclasses of \texttt{LWObservable} should implement the method \texttt{get\_fisher}, which \texttt{DNumberDensityObservable} obtains using eq.~\eqref{lw_ssc} by numerically integrating  eq.~\eqref{number_density_response} as described in subsection~\ref{ssec:obs_response}. Alternatively, to use the more efficient numerical Fisher matrix manipulation method described in subsection~\ref{sssec:converged_lw}, an \texttt{LWObservable} may instead implement a \texttt{get\_perturbing\_vector} method, which for \texttt{DNumberDensityObservable} returns $\frac{\partial \Delta n_{12}}{\partial \delta^\alpha}$ and $(N_{12})^{-1}$ for the perturbing vector ${\bf V}$ and inverse variance ${\bf K}$ respectively. For the current implementation $(N_{12})^{-1}$ is required to be diagonal. Which method should be used is toggled by setting the \texttt{LWObservable} object's \texttt{fisher\_type} variable to \texttt{True} to use \texttt{get\_fisher} and \texttt{False} to use \texttt{get\_perturbing\_vector}.

\subsection{Cosmological Parametrizations}
\label{ssec:cosmo_param}

In order to properly calculate the response of a set of short-wavelength observables $\{O^I\}$ to cosmological parameters $\{\Theta^i\}$, ${\partial O^I}/{\partial \Theta^i}$, we must select a cosmological parametrization. The parametrization is specified in the \texttt{CosmoPie} module, along with a set of rules for calculating derived parameters, so that the parametrization can easily be interchanged. Parameters relating to the dark energy equation of state $w(z)$ are handled separately from the rest. By default, SuperSCRAM uses $\{\Omega_m h^2,\Omega_b h^2, \Omega_{de} h^2, n_s,\ln{A_s}\}$ as its basic set of parameters, where $\Omega_m,\Omega_b,$, and $\Omega_{de}$ are the total matter, baryon, dark energy densities respectively. This parametrization is chosen to be compatible with the parametrization in \cite{jdem_fom}. For simplicity, we set $\Omega_k=0$. If we changed the fiducial model to a curved Universe, then we would have to build the basis modes $\psi_\alpha({\bf r})$ for a curved Universe, using the same ultraspherical Bessel functions that appear in CMB anisotropy calculations \cite{1998ApJ...494..491Z}. We also fix the neutrino masses to be $\sum m_\nu=0$ in the present implementation.

\subsection{Matter Power Spectrum}
\label{ssec:matter_power}
SuperSCRAM provides several interchangeable sample implementations of the matter power spectrum accessible using a \texttt{MatterPowerSpectrum} object, including Python implementations of the Takahashi \cite{takahashi_halofit} and Casarini \cite{casarini_halofit} revisions of the Halofit \cite{smith_halofit} model, linear matter power spectra from CAMB \cite{camb}, and the FAST-PT \cite{fastpt,fastpt2} implementation of the one-loop power spectrum. Additionally, we have extended both Halofit and FAST-PT to facilitate an arbitrary $w(z)$ using the same procedure used in refs.~\cite{casarini_halofit,casarini_halofit_math} to extend the Halofit model as described in Appendix~\ref{app:modify_matter}, although we have not attempted to validate the resulting power spectra through simulations. In this paper, we report only results for the Halofit model.

\subsection{Fisher Matrix Manipulation}
\label{ssec:fisher_objects}

Fisher and covariance matrix manipulations are primarily handled by the \texttt{FisherMatrix} class. The \texttt{FisherMatrix} class is optimized to efficiently perform the manipulations of Fisher matrices required by SuperSCRAM with sizes up to the memory limitations of the machine it runs on. For memory efficiency, it can internally store either a Fisher matrix, covariance matrix, or their Cholesky decompositions. The representations can be exchanged in place internally, and all external class methods intended for use by other modules can be used regardless of the internal representation. For some functions, the choice of internal representation can affect run time significantly, especially for large matrices. 
The overall logic for managing the Fisher matrices for the short and long-wavelength observables and cosmological parameters is handled by the \texttt{MultiFisher} class. A \texttt{MultiFisher} object takes an \texttt{LWBasis} object, an \texttt{SWSurvey} object, and a \texttt{LWSurvey} object. From these objects, it extracts the appropriate projection matrices for converting between the space of long-wavelength and short-wavelength observables, and the space of short-wavelength observables and cosmological parameters, ${\partial O^{I}}/{\partial\delta^\alpha}$ and ${\partial O^i}/{\partial\Theta^i}$ respectively. It then performs the necessary projections and applies priors internally. As an example, if we have already created a Fisher matrix object called \texttt{multi\_f}, we can use
\begin{verbatim}
f_set = multi_f.get_fisher_set(include_priors=False)
\end{verbatim}
to obtain a set of \texttt{FisherMatrix} objects in cosmological parameter space containing just the Gaussian covariance and total covariance with and without mitigation. To then obtain the eigensystems with and without mitigation using the Gaussian covariance as a metric, as described in subsection~\ref{ssec:eig}, we can use
\begin{verbatim}
eig_set = multi_f.get_eig_set(f_set)
\end{verbatim}

\subsubsection{Long-Wavelength Convergence}\label{sssec:converged_lw}

Obtaining well converged results for the super-sample covariance in our formalism may require a long-wavelength Fisher matrix larger than can be stored in memory. However, because the long-wavelength covariance matrix $C^{\alpha\beta}_{\text{SSC}}$ as written in Appendix~\ref{app:basis} is block diagonal, it is possible to calculate the contribution to the short-wavelength covariance matrix $C^{IJ}_{\text{SSC}}$ from each block individually, which requires individual matrices only as large as $(2l_{\text{max}}+1)^2$. However, taking advantage of the block diagonal covariance matrix prevents directly perturbing the Fisher matrix with mitigating information as in eq.~\eqref{F_tot}. Provided it is possible to decompose 
\begin{equation}
F_{\alpha\beta}^{\text{mit}}=({\bf V}^{\rm T} {\bf K V})_{\alpha\beta} = V^a{_\alpha} K_{ab} V^b{_\beta},
\end{equation}
as in eq.~\eqref{lw_ssc},\footnote{The implementation in SuperSCRAM has a diagonal ${\bf K}$, but this is not essential to the formalism.} we may instead use the Woodbury matrix identity 
\begin{equation}\label{woodbury}
C^{\alpha\beta}_{\text{SSC+mit}}=[({\bf C}_{\text{SSC}}^{-1}+{\bf V}^{\rm T} {\bf K} {\bf V})^{-1}]^{\alpha\beta}={\bf C}_{\text{SSC}}^{\alpha\beta}-[{\bf C}_{\text{SSC}}{\bf V}^{\rm T}({\bf K}^{-1}+{\bf VC}_{\text{SSC}}{\bf V}^{\rm T})^{-1}{\bf VC}_{\text{SSC}}]^{\alpha\beta}.
\end{equation}
Here $C^{\alpha\beta}_{\text{SSC+mit}}$ is not necessarily block diagonal, so storing it would require a full-size matrix. However, the matrix
\begin{equation}
W^{aI} \equiv V^a{_\alpha} C_{\rm SSC}^{\alpha\beta} \frac{\partial O^I}{\partial \delta^\beta}
\end{equation}
can be computed without storing the full $C^{\alpha\beta}_{\text{SSC}}$, as can $X^{a\alpha}\equiv V^a{_\beta}L^{\beta\alpha}_{\text{SSC}}$, using the fact that the Cholesky decomposition $L^{\alpha\beta}_{\text{SSC}}$  of a block diagonal matrix $C^{\alpha\beta}_{\text{SSC}}$ is also block diagonal. We can then write
\begin{align}\label{fast_cij}
C^{IJ}_{\text{SSC+mit}}&=\frac{\partial O^I}{\partial \delta^\alpha}C_{\text{SSC}}^{\alpha\beta}\frac{\partial O^J}{\partial \delta^\beta}-
W^{aI} [({\bf K}^{-1} + {\bf XX}^{\rm T})^{-1}]_{ab} W^{bJ},
\end{align}
which can now be computed without ever storing a full-size matrix. In addition to allowing better-converged results due to the much larger number of practically usable basis elements, this procedure is also orders of magnitude faster and is more numerically stable than perturbing the Fisher matrix directly. This procedure is implemented in the \texttt{SphBasisK} module. To allow this procedure to be used, all \texttt{LWObservable} objects must set \texttt{fisher\_type=False} and create a \texttt{get\_perturbing\_vector} method, which returns ${\bf V}$ and ${\bf K}$, for all long-wavelength observables in the survey. We implement this functionality for our \texttt{DNumberDensityObservable} class. 

\subsection{Long-Wavelength Basis}
A basis for long-wavelength observations can be specified as an extension of the \texttt{LWBasis} class. A long-wavelength basis must at minimum implement \texttt{get\_fisher()}, which returns a \texttt{FisherMatrix} object, and \texttt{get\_dO\_I\_ddelta\_alpha(geo,integrand)}, which calculates ${\partial O^I}/{\partial\delta^\alpha}$ by integrating an observable input in a grid $[{\partial O^I}/{\partial \bar{\delta}}](z)$, where the grid should correspond to the fine $z$ grid of a given \texttt{Geo} object. The long-wavelength basis described in Appendix~\ref{app:basis} is provided by the \texttt{SphBasisK} class. 

\section{Applications}\label{sec:applications}
\subsection{Example Application: WFIRST+LSST}\label{ssec:example}

As a demonstration, we calculate the importance of the super-sample covariance in a simulated WFIRST weak lensing survey footprint, with mitigation from a simulated LSST footprint.  For the LSST footprint, we use a simplified survey window from $+5^\circ$ to $-65^\circ$ equatorial latitude spanning $360^\circ$ in longitude, with a $\pm20^\circ$ mask around the galactic plane, which gives about $13600$ deg$^2$ of usable survey area. For the WFIRST footprint, we use an approximately $2100$ deg$^2$ footprint, depicted along with the LSST footprint in the right panel of figure~\ref{fig:sample_geos}. Neither footprint is necessarily what will be implemented; for example, the LSST DESC has proposed an extended footprint going farther north \cite{2018arXiv181200515L}.

For LSST, we use the forecast number density of galaxies for the LSST year 1 dataset with $i<24.1$ from ref.~\cite{lsst_sci_req}, which gives a total of about $18$ galaxies/arcmin$^2$. We choose the $i<24.1$ cutoff to give a conservative estimate of the number of galaxies with good photometric redshifts; by the time of the WFIRST analysis, the photometric redshifts for these bright objects should be very good. We use 4 evenly spaced redshift bins of width $\Delta z=0.3$ from $z=0$ to $z=1.2$, which should be very large compared to the photometric redshift uncertainties. For the lensing source galaxies, we use the WFIRST weak lensing sources, using typical High Latitude Survey parameters of 5 exposures of 140 s each. The source densities were computed using v15 of the WFIRST Exposure Time Calculator \cite{2013ascl.soft11012H} and the Phase A throughput tables.\footnote{See {\tt https://wfirst.gsfc.nasa.gov/science/WFIRST\_Reference\_Information.html}} The input galaxy catalog was from the CANDELS GOODS-S region \cite{2013ApJS..207...24G}, using the photometric redshifts from ref.~\cite{2014ApJ...796...60H}. Cuts were applied requiring detection S/N of the galaxy $>18$, ellipticity error per component $<0.2$, and resolution factor $>0.4$ in the convention of ref.~\cite{2002AJ....123..583B}. These cuts leave about $34$ galaxies/arcmin$^2$ for H band only. The actual source density will be larger in the combined images from all 4 filters, but, for simplicity, we focus only on the H band. The results are most sensitive to the number density chosen for the WFIRST lensing source galaxies because increasing the number of sources decreases the Gaussian covariance. These number densities should give an adequate approximation of the survey galaxy number densities for purposes of this demonstration. 

The intent of this demonstration is to give a rough idea of the impact of super-sample covariance alone on a realistically shaped survey footprint and to give a rough idea of how much improvement could be expected from a simple mitigation strategy. It is not intended to be a full forecast of the actual constraints WFIRST will be able to achieve. In particular, we do {\em not} include any systematic or non-Gaussian effects beyond the super-sample covariance. This is because our philosophy is to compare each added term in the error budget (such as SSC) to the irreducible statistical error bars, and not to the combination of all other uncertainties where we have mitigation efforts in progress.

\begin{figure}[!ht]
 \centering
\includegraphics[width=1\textwidth]{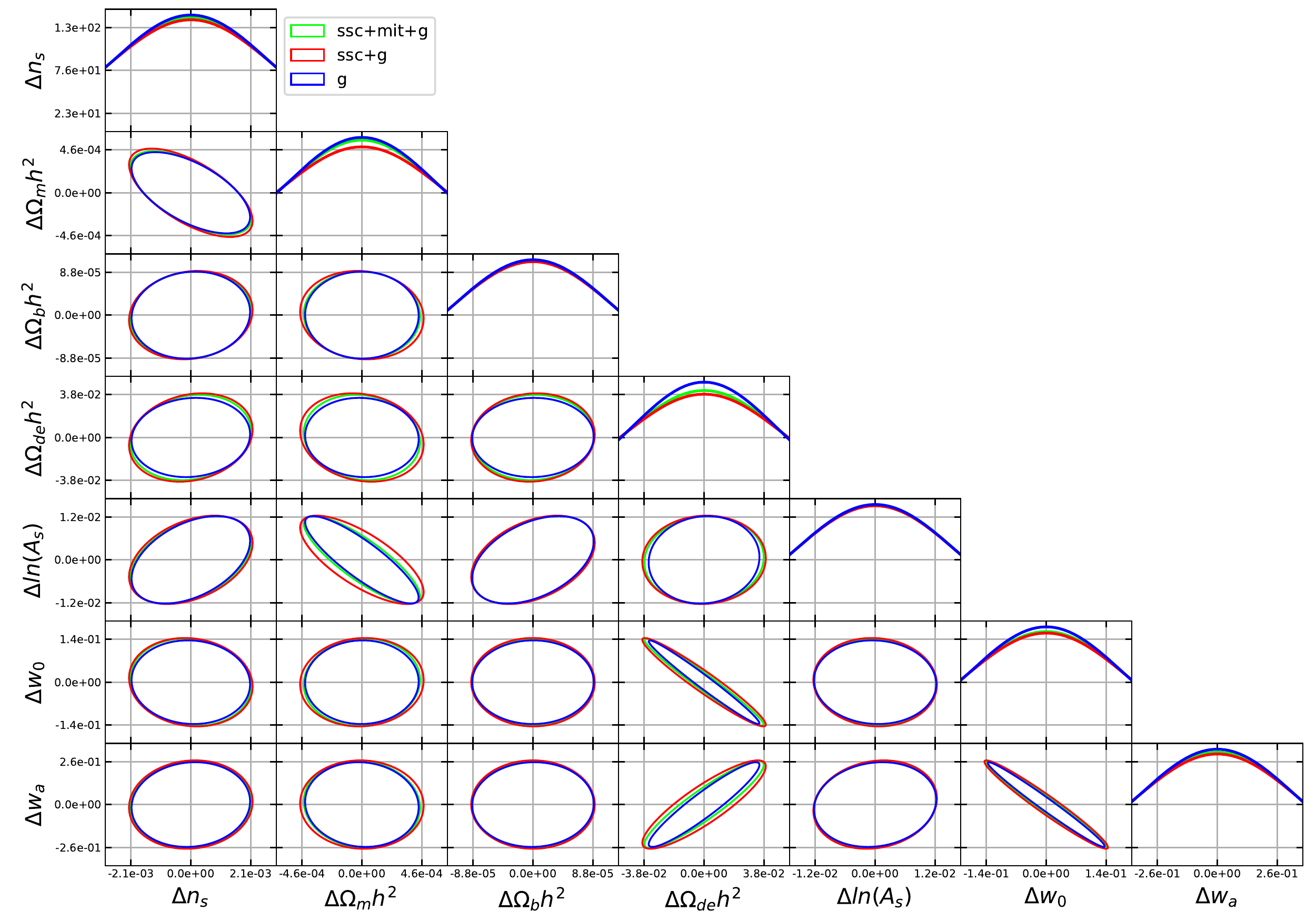}
  \caption{Impact of super-sample covariance and mitigation on cosmological parameter constraints. Constraints in this plot use $\Delta \chi^2=2.3$, with a set of forecast Planck constraints from the WMAP5 fiducial point as cosmological priors included \cite{jdem_fom}. Although it is apparent that super-sample covariance has an effect, the appearance of this plot is very sensitive to the choice of parametrization and the priors applied, as can be seen by comparison to figure~\ref{fig:ellipse_plot_h_1_percent}. Therefore, it is not possible from this plot alone to accurately assess the magnitude of the super-sample effect. Figure~\ref{fig:par_plot} is more useful for isolating the effect of super-sample covariance.}
\label{fig:ellipse_plot}
\end{figure}

\begin{figure}[!ht]
 \centering
\includegraphics[width=1\textwidth]{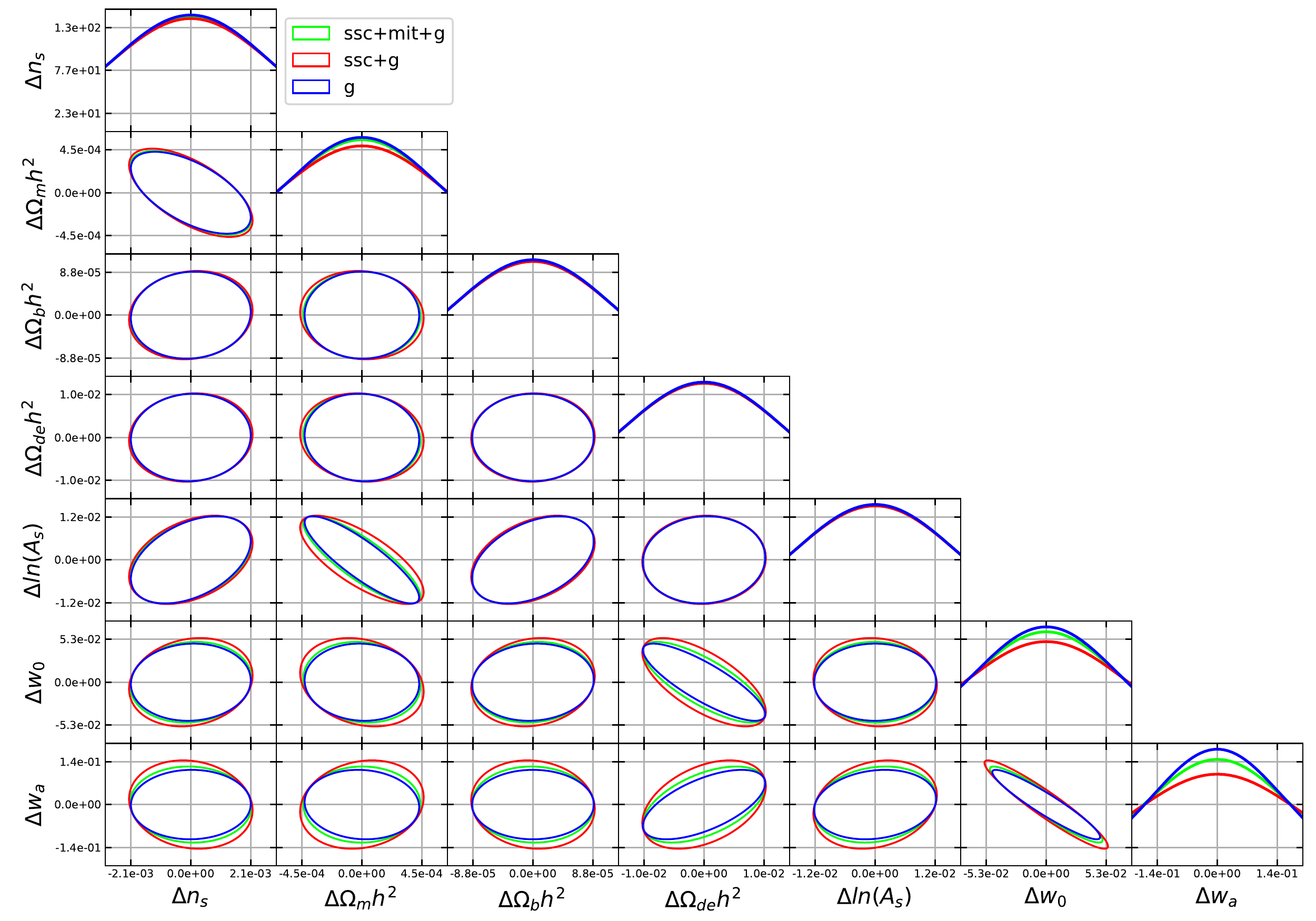}
  \caption{The same run as figure~\ref{fig:ellipse_plot}, but with an additional $1\%$ prior applied to $h$. The apparent effect of super-sample covariance on cosmological parameter space is now significantly different. $\Omega_{\text{de}}h^2$ is now well enough constrained that the super-sample effect is hardly noticeable, while $w_0$ and $w_a$ now appear significantly contaminated by super-sample covariance, where they appeared much less contaminated previously. Because it is impossible to know at this stage the exact set of priors that an experiment like WFIRST will apply, this plot and figure~\ref{fig:ellipse_plot} are both plausible forecasts. Because the conclusions about the impact of super-sample covariance drawn from such plots can vary significantly within the range of reasonable analysis choices, we recommend instead that the super-sample effect be assessed directly using our formalism, as in Table~\ref{tab:eig_tab} and figure~\ref{fig:par_plot}.}
\label{fig:ellipse_plot_h_1_percent}
\end{figure}

\begin{figure}[!ht]
\centering
\includegraphics[width=1\textwidth]{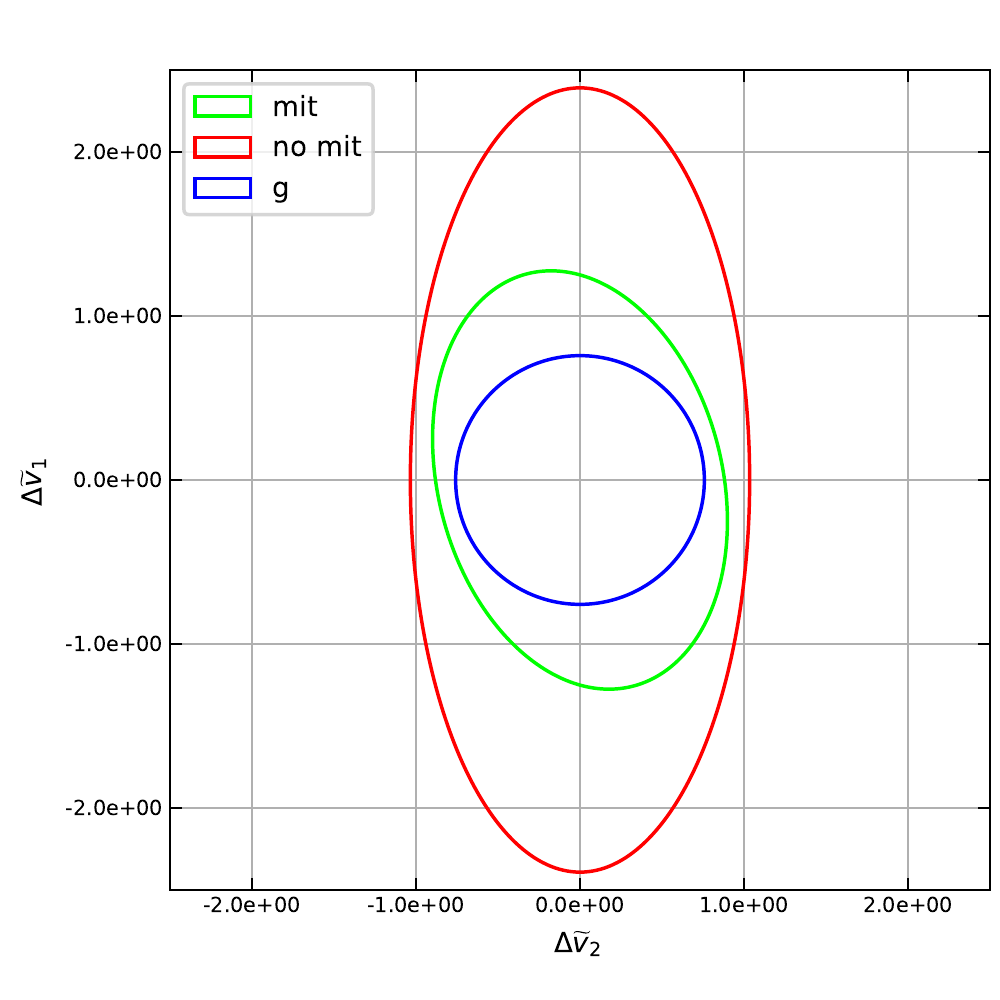}
 \caption{The covariances $\widetilde{C}_{\text{SSC},\text{g}}$, $\widetilde{C}_{\text{SSC},\text{SSC}}$, $\widetilde{C}_{\text{SSC},\text{mit}}$, as defined in eq.~\eqref{oneform_covariance} for the two directions in cosmological parameter space most affected by super-sample covariance. The Gaussian $1-\sigma$ contour is a circle by construction, and the unmitigated super-sample component is diagonal, such that the axes of the `no mit' ellipse align with the axes of the plot. The rotation of the `mit' ellipse due to mitigation shows that the most contaminated direction has changed somewhat after mitigation, which is shown in Table~\ref{tab:eig_tab}. Nearly all of the super-sample effect is concentrated to these 2 combinations of parameters. Note that the magnitude of the contamination is not apparent from figure~\ref{fig:ellipse_plot}. Additionally, this plot is not drastically affected by changes to the parametrization, such as fixing $w_a=0$, while such changes affect figure~\ref{fig:ellipse_plot} significantly.}
\label{fig:par_plot}
\end{figure}

In Figures~\ref{fig:ellipse_plot}~and~\ref{fig:ellipse_plot_h_1_percent}, we show traditional triangle plots of our constraints with different reasonable priors applied. It is apparent from these plots that super-sample covariance has an effect. However, the overall magnitude is difficult to read off, and different choices of priors change the apparent effect of super-sample covariance, which is undesirable for comparison purposes. Instead, we recommend comparing the magnitude of the effect in the most contaminated directions in parameter space, as described in subsection~\ref{ssec:eig}. The comparison is shown in figure~\ref{fig:par_plot}, with the corresponding effect summarized numerically Table~\ref{tab:eig_tab}. From figure~\ref{fig:par_plot}, it is much more apparent that there exist directions significantly affected by super-sample covariance. Additionally, our formalism removes the dependence on the choice of priors by isolating the effect of super-sample covariance on the constraining power of the survey. Figure~\ref{fig:par_plot} and Table~\ref{tab:eig_tab} show more clearly that a mitigation strategy can significantly reduce the amount of constraining power lost due to super-sample covariance. 

\begin{table}[!htbp]\begin{center}
\begin{tabular}{llllllllll}
&$\Pi_{(i)}\lambda^{(i)}$&$\lambda^{\text{top}}$&$\widehat{n_s}$&$\widehat{\Omega_m h^2}$&$\widehat{\Omega_b h^2}$&$\widehat{\Omega_{\text{de}}h^2}$&$\widehat{ln(A_s)}$&$\widehat{w_0}$&$\widehat{w_a}$\\
$C_{\text{SSC}}$&20.26&9.94&329.3&5051.1&-8425.4&-390.9&269.4&-206.3&-58.6\\
$C_{\text{mit}}$&4.02&2.93&271.0&4319.2&-7132.2&-346.2&229.2&-176.5&-48.9\\
\end{tabular}
\caption{Results from a well converged run of SuperSCRAM with a sample WFIRST and LSST survey footprints. with $k_\text{cut}=0.0814\;\text{Mpc}^{-1}$ as described in Appendix~\ref{app:basis}, with the product of eigenvalues, eigenvalue in most contaminated direction, and the coefficients describing the most contaminated directions, as described in subsection~\ref{ssec:eig}. The convergence of $\Pi_{(i)}\lambda^{(i)}$ with respect to $k_\text{cut}$ is $\lesssim\mathcal{O}(1\%)$. The mitigation reduces the volume of parameter space contaminated by SSC by approximately a factor of 5. The most contaminated direction does not change drastically after mitigation.}
\label{tab:eig_tab} 
\end{center}
\end{table}

\subsection{Sensitivity to Survey Geometry}\label{ssec:survey_geometry}
\begin{figure}[!ht]
 \centering
\includegraphics[width=1\textwidth]{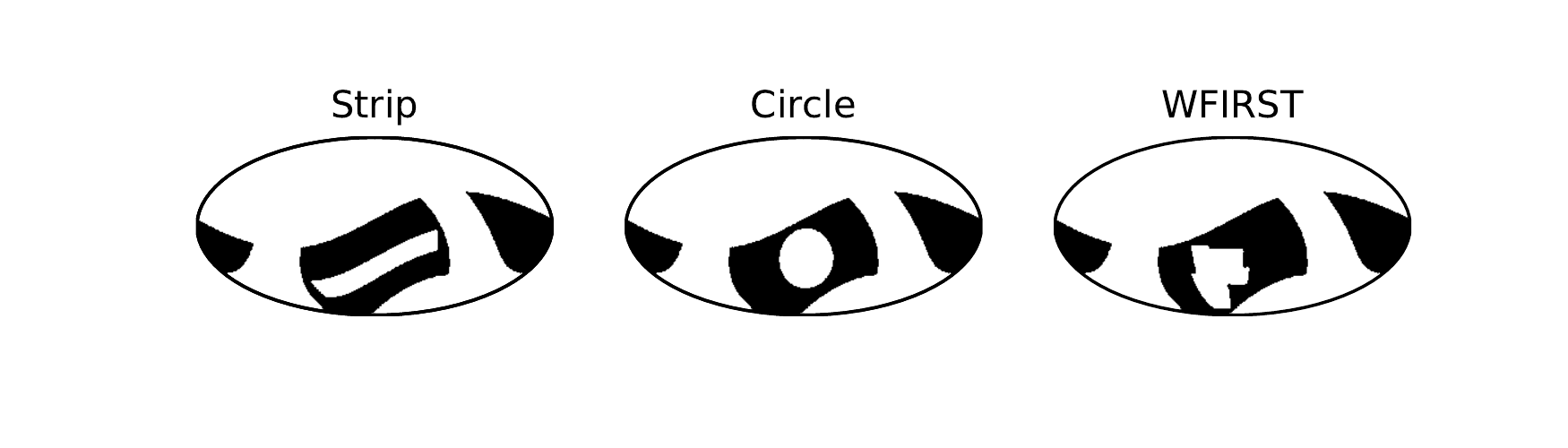}
  \caption{Three example geometries tested. The black region is the LSST-like survey window, with white cutouts showing three different possible geometries for the WFIRST-like survey with identical angular areas $2098.2\;\text{deg}^2$. With all parameters other than the angular window the same as the run summarized in Table~\ref{tab:eig_tab}, we find, from left to right, $\Pi\lambda_{\text{g},SSC}=21.98,20.20,20.26$, and $\Pi\lambda_{\text{g},mit}=3.92,4.05,4.02$. Without mitigation, the results distinctly favor more compact survey geometries. Mitigation may be more effective for extended geometries, such that the overall constraint with mitigation is slightly better for the "Strip" geometry, although it should be noted that our mitigation strategy is primarily for demonstration purposes and cannot be expected to precisely reflect the actual performance of a realistic mitigation strategy. The more significant result is that, provided WFIRST is completely embedded in and LSST-like survey window, the specific shape of the WFIRST survey window is largely a secondary consideration for the purposes of mitigating super-sample covariance.}
\label{fig:sample_geos}
\end{figure}

One application of SuperSCRAM is to test the impact of varying the survey geometry. Previous analytic studies \cite{krause_eifler_cosmolike} have generally considered only a circular survey geometry. Results comparing several different survey geometries are shown in figure~\ref{fig:sample_geos}. The magnitude of the effect varies at approximately the $O(10\%)$ level between different contiguous, reasonably compact high latitude survey geometries. Because more compact geometries produce a smaller super-sample covariance, estimates using only a circular geometry may tend to understate the magnitude of the effect. The differences between geometries are small enough that it is possible that including additional super-sample effects beyond the density contrast effect considered here, such as tidal effects and redshift-space distortions, could change the relative ordering of the effect on different survey geometries. Because the mitigation reduces the super-sample contamination far more than the differences between any reasonable geometries, the possibility of applying mitigation strategies appears to be a more important survey-design consideration than the detailed survey geometry for controlling super-sample covariance. 

\subsection{Other Applications}\label{ssec:extensions}

A variety of possible applications are would be straightforward to implement but are not currently included. Additional short-wavelength observables, such as galaxy-galaxy lensing and galaxy power spectra, could be added to better understand the full constraining power of a weak lensing survey. 

Many more mitigation strategies are possible. The current mean number density observable is only a simple demonstration of the information that can be extracted from a larger survey and does not include information about galaxy number density as function of sky position. Additionally, CMB lensing could be useful to constrain over-densities at higher redshifts. The mean tangential shear could also provide additional information.

It would be useful to compare the results of our Fisher matrix method to a Monte Carlo code. To achieve more realistic forecasts, photometric redshift uncertainties and other systematic and non-Gaussian effects would need to be implemented. 

SuperSCRAM can be used to investigate the contribution of super-sample covariance to the tradeoff between a wider and a deeper survey. A more detailed model of the survey's sensitivity as a function of redshift would be necessary to draw useful conclusions from this possible application. 

\section{Conclusions}\label{sec:conclusion}

In this paper, we have developed a Fisher matrix formalism and present the code SuperSCRAM for evaluating the impact of super-sample covariance on a realistically shaped weak lensing survey geometry. We have further applied our formalism to investigate the possibility of obtaining information from wider, shallower surveys to reduce the impact of super-sample covariance. Overall, we find that mitigation strategies can improve constraining power enough to merit serious consideration for next-generation weak lensing surveys. 

Further work is needed to investigate other contributions to super-sample covariance beyond the density contrast effect discussed in this paper, such as tidal effects and redshift-space distortions. Including those effects would improve the ability to make detailed conclusions about the relative effectiveness of different survey geometries. 

Additionally, it would be productive to investigate a wider array of mitigation strategies. In the simplest case, our mean number density observable does not exhaust the information available from a larger wider survey, as a more detailed analysis including observed number density as a function of sky position would likely improve that observable. Additionally, CMB lensing could potentially provide information about densities at much higher redshifts than can be obtained by a shallow galaxy survey, which could significantly improve constraining power. It may also be possible to use the tangential shear and other weak lensing related observables as further mitigating observables.

\acknowledgments
 
M.C.D. and C.M.H. were supported by the Simons Foundation, NASA, and the US Department of Energy during the preparation of this work.

We thank Elisabeth Krause for providing scripts to run CosmoLike for calibration of SuperSCRAM, as well as for general discussions. We thank Tim Eifler, Olivier Dor\'e, Xiao Fang, and Benjamin Buckman for useful feedback.

The computations in this paper were run on the CCAPP condos of the Ruby and Pitzer Clusters at the Ohio Supercomputer Center.

This work is based in part on observations taken by the CANDELS Multi-Cycle Treasury Program with the NASA/ESA HST, which is operated by the Association of Universities for Research in Astronomy, Inc., under NASA contract NAS5-26555.

\bibliographystyle{JHEP.bst}
\bibliography{main}

\appendix 

\section{Basis Decomposition}
\label{app:basis}

In SuperSCRAM, we use a form of a spherical Fourier-Bessel basis, similar to the one described in ref.~\cite{spherical_fourier_bessel}. We use a spherical Bessel function for our radial basis and the real spherical harmonics for our angular basis, so that the elements of our basis can be written:
\begin{align} \label{basis_decomp}
\psi_\alpha(r, \theta, \phi) = j_{l_\alpha}(k_\alpha r) Y^\text{R}_{l_\alpha m_\alpha}(\theta, \phi) ~, 
\end{align} 
where $Y^\text{R}_{lm}(\theta, \phi) $ is the real spherical harmonic, $k_\alpha$ is a solution of $j_{l_\alpha}(k_\alpha R_{max})=0$, and $\alpha$ is an index running over all possible sets $(k_\alpha,l_\alpha,m_\alpha)$. To evaluate the covariance matrix with a finite number of modes, we choose a cutoff $k_{\text{max}}$ and take all combinations of $(k_\alpha,l_\alpha,m_\alpha)$ which have $k_\alpha<k_{\text{max}}$. This cutoff procedure has the effect of gradually decreasing the number of modes included for a given $l_\alpha$ as $l_\alpha$ gets larger, and should converge faster than an arbitrary $l_{\text{max}}$ cutoff. The real spherical harmonics can be defined in terms of Legendre polynomials:
\begin{align}\label{real_harmonic2}
Y^\text{R}_{lm} = \sqrt{\frac{2(l+1)}{4\pi}}\sqrt{\frac{(l-|m|)!}{(l+|m|)!}}(-1)^{|m|}P^{|m|}_l(\cos{\theta})
\begin{cases}
\sqrt{2}\sin{|m|\phi}&\text{if } m<0\\
1& \text{if } m=0\\
\sqrt{2}\cos{|m|\phi}&\text{if } m>0
\end{cases}\,.
\end{align}
We can then write a density fluctuation mode in this basis as
\begin{align}
\delta^\alpha(k_\alpha)& = \frac{1}{N_\alpha}\int \delta(\mathbf{r}) \jl{\alpha}(k_\alpha r) \sph{\alpha}{r} d^3 \mathbf{r}~\nonumber\\
& = \frac{1}{N_\alpha}\int \dk \delta(\mathbf{k}) \int_0^{r_\text{max}} dr r^2 \jl{\alpha}(k_\alpha r) \int d^2 \hat{\bf r}\, e^{ikr \hat{\bf k} \cdot \hat{\bf r}} \sph{\alpha}{\bf r} \nonumber\\
& =\frac{4 \pi i^{l_\alpha}}{N_\alpha}\int  \dk \delta(\mathbf{k})  \sph{\alpha}{k}  \int_0^{r_\text{max}} dr r^2 \jl{\alpha}(k_\alpha r) \jl{\alpha}(kr)  ~,
\label{ss_mode}
\end{align}
where $N_\alpha$ is a normalization factor. In the second equality we use the Fourier transform relation $\delta(\mathbf{r})=(2\pi)^{-3} \int d^3 \mathbf{k}\, e^{i \mathbf{k} \cdot \mathbf{r}} \delta(\mathbf{k})$, and in the third equality we use the identity
\begin{align} 
\int_{S^2} d^2 \hat{\bf r}\, \sph{\alpha}{r} e^{i \mathbf{k} \cdot \mathbf{r}} = 4 \pi i^{l_\alpha} j_{l_\alpha}(k r)\sph{\alpha}{k}~. 
\end{align}
The normalization factor $N_\alpha$ is defined
\begin{align}
N_\alpha&=\int{d^3\mathbf{r}\,\sph{\alpha}{r} \sph{\alpha}{r} j_\alpha(k_\alpha r) j_\alpha(k_\alpha r)} \nonumber\\
&=  I_\alpha(k_\alpha, r_{\text{max}})  \int_\Omega \sph{\alpha}{r} \sph{\alpha}{r}\, d^2\hat{\bf r} \nonumber\\
& =I_\alpha(k_\alpha, r_{\text{max}}),
\label{normalization}
\end{align}
where $ \int_\Omega \sph{\alpha}{r} \sph{\alpha}{r} \,d^2\hat{\bf r}=1$. The integral $ I_\alpha(k_\alpha, r_{\text{max}}) $ is defined as
\begin{align}
\label{i_alpha}
\begin{split} 
I_\alpha(k,  r_\text{max}) & =  \int_0^{r_\text{max}}dr\, r^2 \jl{\alpha}(k_\alpha r) \jl{\alpha}( k r) \\
& = \frac{\pi}{2} \sqrt{ \frac{1}{k _ \alpha k}} \int_0^{r_\text{max}}dr\, r J_{l_\alpha+ 1/2}(k_\alpha r)  J_{l_\alpha+ 1/2}( k r) \\
& = \frac{\pi}{2} \frac{ r_\text{max}}{ \sqrt{ k _ \alpha k}}\frac{\left[k_\alpha J_{l_\alpha+ 1/2}( k  r_\text{max})J_{l_\alpha+ 1/2}'(k_\alpha  r_\text{max}) - k J_{l_\alpha+ 1/2}(k_\alpha  r_\text{max})J_{l_\alpha+ 1/2}'(k  r_\text{max})\right]}{k^2 - k_\alpha^2}\\
& = \frac{\pi}{2} \frac{ r_\text{max}}{ \sqrt{ k _ \alpha k}}\frac{\left[k_\alpha J_{l_\alpha+ 1/2}( k  r_\text{max})J_{l_\alpha- 1/2}(k_\alpha  r_\text{max}) - k J_{l_\alpha+ 1/2}(k_\alpha  r_\text{max})J_{l_\alpha- 1/2}(k  r_\text{max})\right]}{k^2 - k_\alpha^2}\\
& = \frac{\pi}{2} \frac{ r_\text{max}}{ \sqrt{ k _ \alpha k}}\frac{k_\alpha J_{l_\alpha+ 1/2}( k  r_{\text{max}})J_{l_\alpha- 1/2}(k_\alpha  r_\text{max})}{k^2 - k_\alpha^2}~,
\end{split} 
\end{align}
where in the last step we have use $J_{l_\alpha+1/2}(k_\alpha r_{\rm max})=0$, from the defining property of $k_\alpha$. In the special case $k=k_\alpha$, we can simplify $N_\alpha = I_\alpha(k,r_\text{max})$ using Bessel function identities:
\begin{align}
N_\alpha = I_\alpha(k_\alpha,  r_\text{max}) & = \lim_{k\to k_\alpha} \frac{\pi}{2} \frac{ r_\text{max}}{ \sqrt{ k _ \alpha k}}\frac{k_\alpha J_{l_\alpha+ 1/2}( k  r_\text{max})J_{l_\alpha- 1/2}(k_\alpha  r_\text{max})}{k^2 - k_\alpha^2}\nonumber\\
&=\lim_{k\to k_\alpha}\frac{\pi r_\text{max}}{2} \frac{J_{l_\alpha+ 1/2}( r_\text{max})J_{l_\alpha- 1/2}(k_\alpha  r_\text{max})}{k^2 - k_\alpha^2}
\nonumber \\
&=\lim_{k\to k_\alpha}\frac{\pi r_\text{max}}{2} \frac{r_\text{max}J'_{l_\alpha+ 1/2}(k r_\text{max})J_{l_\alpha- 1/2}(k_\alpha  r_\text{max})}{2 k}
\nonumber \\
&=\lim_{k\to k_\alpha}\frac{\pi r_\text{max}^2}{2} \frac{\left[\frac{l_\alpha+1/2}{k r_{\text{max}}}J_{l_\alpha+1/2}(k r_\text{max})-J_{l_\alpha+3/2}(k_\alpha{r_\text{max}})\right]J_{l_\alpha- 1/2}(k_\alpha  r_\text{max})}{2 k}
\nonumber \\
&=-\frac{\pi r_\text{max}^2}{4 k_\alpha}J_{l_\alpha+3/2}(k_\alpha{r_\text{max}})J_{l_\alpha- 1/2}(k_\alpha  r_\text{max})\label{norm_final}.
\end{align}
With these definitions, we can calculate the covariance matrix $C^{\alpha\beta}_\text{SSC}=\left(F^{\alpha\beta}_\text{SSC}\right)^{-1}$ in our basis:
\begin{align} 
\begin{split} 
C^{\alpha\beta}_{\text{SSC}}=\langle \delta^\alpha \delta^\beta \rangle &= \frac{ (4 \pi)^2}{N_\alpha N_\beta} \int  \dkv{1} \dkv{2} \,\langle \delta(\mathbf{k}_1) \delta^*(\mathbf{k}_2) \rangle \sph{\alpha}{k_1} \sph{\beta}{k_2} \\
& \;\;\; \times   \int_0^{r_\text{max}}  dr\, r^2 \jl{\alpha}(k_\alpha r) \jl{\alpha}( k_1 r) \int_0^{r_\text{max}}  dr\, r^2 \jl{\beta}(k_\beta r) \jl{\beta}( k_2 r)  \\
&= \frac{ (4 \pi)^2}{N_\alpha N_\beta} \int \dk\, P^{\delta\delta}(k) I_\alpha(k, r_\text{max}) I_\beta(k, r_\text{max})\sph{\alpha}{k_1}Y^R_{l_\beta m_\beta}(-\hat{k_1})\\
&=\frac{2}{\pi N_\alpha N_\beta}\int{k^2\, dk P^{\delta\delta}(k) \delta_{l_\alpha,l_\beta}\delta_{m_\alpha,m_\beta} I_\alpha(k, r_\text{max}) I_\beta(k, r_\text{max})}~,
\label{C_SSC}\end{split} 
\end{align} 
where $P^{\delta\delta}(k)$ is the matter power spectrum, and we used $(2 \pi)^3 \delta^3_\text{D}( \mathbf{k} + \mathbf{k}') P^{\delta\delta}(k) =  \langle \delta(\mathbf{k}) \delta^*(\mathbf{k}') \rangle$.

\subsection{Response of Observables to Density Fluctuations}
\label{ssec:obs_response}

To calculate the response of an observable $O^I$ to a density fluctuation mode $\delta^\alpha$, we can use the chain rule:
\begin{align}\label{chain_obs}
\frac{\partial O^I}{\partial\delta^\alpha}=\int_0^{z_{\text{max}}}dz~\frac{\partial O^I}{\partial \bar{\delta}}(z)\frac{\partial \bar{\delta}}{\partial \delta^\alpha}(z),
\end{align}
provided $[{\partial{O^I}}/{\partial \bar{\delta}}](z)$ can be calculated. In SuperSCRAM, the integral in eq.~\eqref{chain_obs} is accomplished by calculating the integrand on a grid of $z$ values $\{z_i\}$ and using the trapezoidal rule. To calculate $[\partial\bar{\delta}/\partial\delta^\alpha](z_i)$, we expand the mean density fluctuation $\bar{\delta}(z_i)$, in our basis: 
\label{sec:density_fluct}
\begin{align}\label{bar_delta}
\bar{\delta}(z_i)= \displaystyle \sum_\alpha \frac{3}{r_{i+1}^3 - r_{i}^3} \int_{r_i}^ {r_{i+1} }dr ~ r^2 j_{l_\alpha}(k_\alpha r) \delta_\alpha(k_\alpha) \frac{1}{2\sqrt{\pi} a_{00}} \underbrace{ \iint\limits_\Omega d\Omega ~\sph{\alpha}{r}}_{a_{l_\alpha m_\alpha}} ~,
\end{align}
where $r$ is the comoving distance in the range $r_{i}\le r<r_{i+1}$, and $a_{l_\alpha m_\alpha}$ are the real spherical harmonic coefficients of a given survey window function $\Omega$. 
We can use eq.~\eqref{bar_delta} to write the derivative:
\begin{align}
\frac{\partial \bar{\delta} }{ \partial \delta^\alpha}(z_i)=
\frac{3}{r_{i+1}^3 - r_{i}^3} \int_{r_{i}}^ {r_{i+1} }dr ~ r^2 j_{l_\alpha}(k_\alpha r)  \frac{1}{2\sqrt{\pi} a_{00}} a_{l_\alpha m_\alpha}~.
\end{align}

\section{Projected Power Spectra}\label{app:projected_power}

The angular correlation function $w_{AB}(\hat{n}\cdot\hat{n}')$ of the line of sight projections of two fields $A$ and $B$ can be expanded in terms of its angular power spectrum $C_{AB}\left(\ell\right)$ \cite{extended_limber}:
\begin{equation}
w^{AB}(\hat{n}\cdot\hat{n}')\equiv\left<A(\hat{n})B(\hat{n}')\right>=\sum_{\ell}{\frac{2\ell+1}{4\pi}C^{AB}(\ell)P_\ell(\hat{n}\cdot\hat{n}')},
\end{equation}
where $\hat{\bf n}$ and $\hat{\bf n}'$ are unit vectors in the direction of observation and $P_\ell$ are  Legendre polynomials. For a given field $A$, there is a weight function $q^A$ which relates the field to its line of sight projection $\tilde{A}(\hat{n})$, such that \cite{extended_limber}:
\begin{equation}
\tilde{A}(\hat{\bf n}) = \int{dr\, q^A(r)A(r\hat{\bf n})},
\end{equation}
where $r$ is the comoving coordinate. In terms of $q_A$ and $q_B$, the angular cross-power spectrum can be written:
\begin{equation}\label{no limber}
C^{AB}(\ell)\equiv\left<\tilde{A}_{lm}\tilde{B}_{lm}^*\right>=\int{dr_1 dr_2 q_A(r_1)q_B(r_2)\int{\frac{2 k^2 dk}{\pi} j_\ell(k r_1)j_\ell(k r_2)P^{AB}(k)}},
\end{equation}
where $j_\ell(kr)$ are spherical Bessel functions, and $k=(\ell+\frac{1}{2})/{r}$. In practice, this integral is inconvenient to compute numerically due to the rapid oscillations of the spherical Bessel functions for $kr\gtrsim \ell+\frac{1}{2}$. Therefore, most authors write $C^{AB}$ using the Limber approximation, which can be taken by expanding this expression to lowest order in $(\ell+\frac{1}{2})^{-1}\ll 1$,
\begin{equation}\label{limber_cab}
C^{AB}(\ell)\cong\int_{0}^{r_{max}}{dr\, \frac{q^A(r)q^B(r)}{r^2}P^{\delta\delta}\left(k=\frac{\ell+1/2}{r}\right)}.
\end{equation}

SuperSCRAM currently implements eq.~\eqref{limber_cab} in the \texttt{ShearPower} class. Note that for $\ell\gg\frac{1}{2}$ many authors take $\ell+\frac{1}{2}\approx\ell$, although SuperSCRAM does not, both because the correction does not affect the code's execution time, and for consistency with other implementations such as the one in CosmoSIS \cite{cosmosis}. The next order correction is suppressed by a factor of $\mathcal{O}((l+\frac{1}{2})^{-2})$, which should be negligible for next generation weak lensing surveys. 

For the weight functions $q_A(r)$ for shear-shear lensing, implemented as \texttt{QShear}, we use \cite{eifler_krause_cosmolike,krause_eifler_cosmolike} 
\begin{equation}\label{shear_weight}
q_{\gamma i}(r) = \frac{3H_0^2\Omega_{m}}{2c^2}\frac{r}{a}g_i(r)
\end{equation}
where
\begin{equation}\label{shear_g}
g_i(r)=\int^{r_{i}+1}_{r_i}p_i(r')(r'-r)/r' dr'
\end{equation}
where $p_i(r')$ is the probability density function for source galaxies in tomographic bin $i$. 
\section{Analytic Polygon Geometry}
\label{app_analy_poly}

Using Stokes's theorem, for a spherical polygon survey window, which has $N$ sides which are great circle arcs, we can write 
\begin{equation}\label{stokes_alm}
a_{l m}=\iint\limits_\Omega d\Omega ~Y^R_{l m}(\hat{r})
 =\sum_{n=1}^{N}\underbrace{\frac{1}{l(l+1)}\int_{\partial\Omega_n}\left[{\boldsymbol\nabla}Y^R_{l m}(\bf\hat{r})\right]\cdot {\bf{\hat{z}}}_n\, ds}_{\equiv\Delta a_{l m,n}},
\end{equation}
where $\partial \Omega_n$ denotes integration over the boundary of the $n$th arc and $\hat{z}_n$ is a unit vector orthogonal to the two unit vectors whose tips touch the ends of the $n$th arc, such that if ${\bf\hat{p}}_{n}$ is the unit vector at the start of the $n$th arc, ${\bf\hat{z}}_n\equiv{{\bf\hat{p}}_{n+1} \times {\bf\hat{p}}_{n}}/{|{\bf\hat{p}}_{n+1} \times {\bf\hat{p}}_n|}$, leaving ${\bf\hat{y}}_n\equiv {\bf\hat{z}}_n \times {\bf\hat{p}}_{n}$. The integral is most simple to evaluate if the great circle is along the equator at $\theta=\frac{\pi}{2}$, so for each side we rotate to a coordinate system where the side is along the equator, calculate $\Delta a'_{l m,n}$ and rotate back to the global coordinate system using 
\begin{align}\label{harmonic_rotate}
\Delta a_{l m,n}=\sum_{m'=-l}^{m'=l}{D^n_{l m m'}\Delta a'_{l m',n}},
\end{align}
where $D^n_{l m m'}$ is a spherical harmonic rotation matrix element. 

The integrand restricted to the equator $\theta=\frac{\pi}{2}$ is given. Recalling eq.~\eqref{real_harmonic2}, we have
\begin{align}\label{y_grad_simpl}
\left[{\boldsymbol\nabla}Y^R_{l m}(\bf\hat{r})\right]\cdot {\bf\hat{z}}_n&\bigg\vert_{\theta=\pi/2}=\frac{\partial Y^R_{l m}(\theta,\phi)}{\partial z}\bigg\vert_{\theta=\pi/2}\nonumber\\
&=-\sin\theta\frac{\partial Y^R_{l m}(\hat{r})}{\partial\theta}\bigg\vert_{\theta=\pi/2}\\&=
-(l+|m|) (-1)^{|m|}\sqrt{\frac{(2l+1)}{4\pi}\frac{(l-|m|)!}{(l+|m|)}}
P^{|m|}_{l-1}(0)\begin{cases}
\sqrt{2}\cos(|m|\phi) &m>0\\
1&m=0\\
\sqrt{2}\sin(|m|\phi) &m<0
\end{cases}\nonumber
\end{align}
where in the last step we have have used the recurrence relation 
\begin{equation}\label{spherical_harmonic_recurrence}
(1-x^2)\frac{d P^m_l(x)}{dx}=l x P^m_l(x)-(l+m) P^m_{l-1}(x).
\end{equation}

Then, because the side has been rotated along the equator, we need only integrate $\phi$ from 0 to the side length $\beta_n$ and find
\begin{align}
\Delta a'_{l m,n} &=\frac{1}{l(l+1)}\int_0^\beta\left[{\boldsymbol\nabla}Y^R_{l m}(\bf\hat{r})\right]d\phi\label{harmonic_side}\\
&=\frac{(-1)^{m}}{l(l+1)}\sqrt{\frac{2l+1}{2l-1}}\sqrt{(l-|m|)(l+|m|)}Y^R_{(l-1) m}\left(\frac{\pi}{2},0\right)
\begin{cases}
\frac{1}{|m|}\sin(\beta_n |m|) &m>0\\
\beta_n &m=0\\
\frac{1}{|m|}(1-\cos(\beta_n |m|)) &m<0
\end{cases}\nonumber
\end{align}
where we have substituted $P_{(l-1)}^m(0)$ for $Y^R_{(l-1) m}(\frac{\pi}{2},0)$ using  eq.~\eqref{real_harmonic2}.
The side length is given $\beta_n\equiv\text{atan2}\left[|{\bf\hat{p}}_{n+1}\times{\bf\hat{p}}_n|,{\bf\hat{p}}_{n+1}\cdot{\bf\hat{p}}_{n}\right]$.

\begin{figure}[!ht]
 \centering
\includegraphics[width=0.5\textwidth]{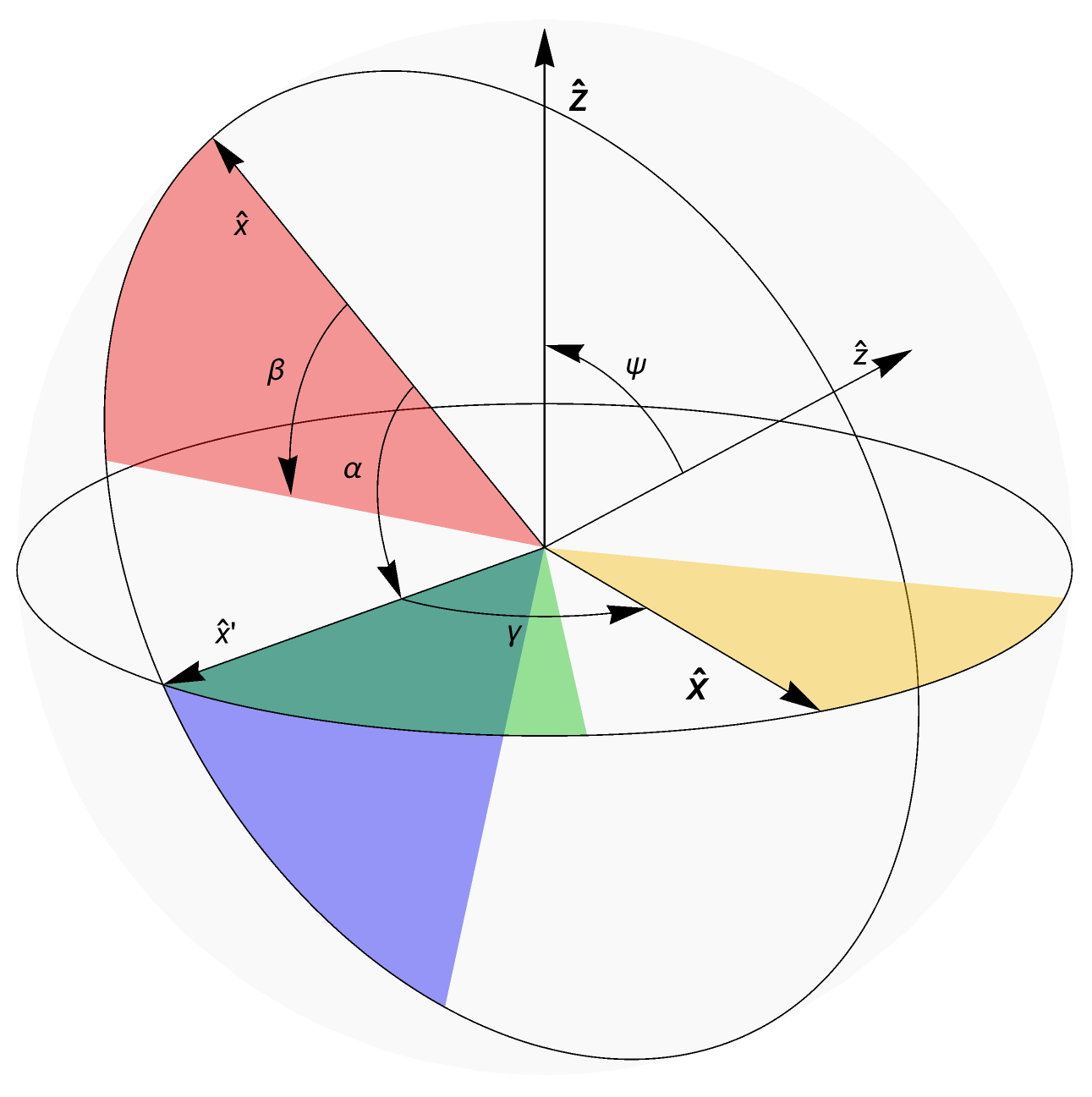}
  \caption{Applying Euler rotations to rotate an arc of angle $\beta$, highlighted in red, to a coordinate system where the side is along the equator, highlighted in yellow. The contribution $\Delta a'_{lm,n}$ is simple to calculate along the equator. Then the contribution to the geometry on the sky $\Delta a_{lm,n}$ may be obtained by applying the rotations to the matrix $\Delta a'_{lm,n}$.}
\label{fig:euler_rotation}
\end{figure}
Now, to find the spherical harmonic rotation matrices, we obtain the Euler angles $\alpha_n$, $\psi_n$, and $\gamma_n$ for the $z-x-z$ Euler rotation necessary to rotate an arc from the global frame into a frame where the side lies along the equator in the $x-y$ plane. as depicted in figure~\ref{fig:euler_rotation}. We define $\psi_n\equiv-\cos^{-1}\left({\bf\hat{z}}_{n z}\right)$, $\gamma_n\equiv \text{atan2}\left[{-{\bf\hat{z}}_{n x}},{{\bf\hat{z}}_{n y}}\right],$ $\alpha_n\equiv-\text{atan2}\left[-{\bf\hat{y}}_{n z},{\bf\hat{x}}_{n z}\right]$. Here  are, by construction, the Euler angles for a $z-x-z$ Euler rotation which transforms from a local coordinate system where the $n$th arc lies in the $x-y$ plane to the global coordinate system, so that 
\begin{align}\label{composite_rotation}
D^n_{l m m'}=\sum_{m_1,m_2=-l}^{l}{D^z_{l m m_1}(\gamma_n)D^x_{l m_1 m_2}(\psi_n)D^z_{l m_2,m'}}(\alpha_n),
\end{align}
where the $z$-rotation matrices are given such that:
\begin{align}\label{rot_z}
\sum_{m'=-l}^l{D^z_{l m m'}(\omega) a'_{l m'}= 
\begin{cases}
\cos(|m|\alpha)a'_{l |m|}+\sin(|m|\alpha)a'_{l -|m|}& m>0\\
a'_{l |m|}& m=0\\
-\sin(|m|\alpha)a'_{l |m|}+\cos(|m|\alpha)a'_{l -|m|}& m<0
\end{cases}}.
\end{align}
We use an angle-doubling algorithm to compute the matrix ${\bf D}^x_l$. Writing ${\bf D}^x_{l}$ as a $(2l+1)\times (2l+1)$ matrix, ${\bf D}^x_{l}(\psi)={\bf E}_{l}(\psi)+\mathbb{1}$, and ${\bf E}_{l}(\psi)$ is computed by recursively applying the angle doubling formula ${\bf E}_{l} = 2 {\bf E}_{l}(\frac{\psi}{2})+[{\bf E}_{l}(\frac{\psi}{2})]^2$ to the infinitesimal rotation matrix ${\bf E}_{l}$. This matrix is constructed as ${\bf E}_{l}(\epsilon)=\mathcal{M}_l^{-1}\tilde{\bf E}_{l}\mathcal{M}_l$, where the infinitesimal complex-basis rotation around $x$ is
\begin{align}\label{E_complex}
\tilde{E}_{lmm'}(\epsilon)&=-i\epsilon [L_x]_{mm'}\nonumber\\
&=\frac{-i\epsilon}{2}\left[\delta_{m,m'+1}\sqrt{(l+1-m)(l+m)}+\delta_{m,m'-1}\sqrt{(l-m)(l+1+m)}\right],
\end{align}
and $\mathcal{M}_l$ transforms to the real spherical harmonic basis:
\begin{align}\label{m_mat}
\mathcal{M}_{l m m'}=
\begin{cases}
\frac{1}{\sqrt{2}}&\text{if } m>0\text{ and } m'=m\\
-i\frac{1}{\sqrt{2}}&\text{if } m>0\text{ and } m'=-m\\
\frac{(-1)^m}{\sqrt{2}}&\text{if } m<0\text{ and } m'=-m\\
i\frac{(-1)^m}{\sqrt{2}}&\text{if } m<0\text{ and } m'=m\\
1&\text{if } m=0\text{ and } m'=0\\
0&\text{otherwise}.
\end{cases}
\end{align}
In SuperSCRAM, the number of doublings can be specified by the user; for our WFIRST+LSST demo, 31 doublings are sufficient to recover $a_{lm}$ up to $l=100$ to within a maximum error of $\lesssim0.1\%$, and approximately 82 doublings will converge all elements of the rotation matrices up to $l=200$ to within the limits of IEEE 745 64 bit precision. The repeated multiplications of $(2l+1)\times (2l+1)$ matrices are relatively time consuming for large $l$. An advantage of the analytic solution over a pixel based geometries is that the results are less vulnerable to pathological input survey geometries, such as a geometry with many narrow stripes (perhaps chip gaps) masked out, aligned so that none of the stripes contained any pixel centroids, which could produce poorer quality results in a pixel based geometry. Additionally the analytic solution can in principle be calculated for arbitrarily large $l$, while the pixelated solution is limited by the resolution of the pixelation scheme. 

The various manipulations required to calculate eq.~\eqref{harmonic_rotate} are computed by the \texttt{alm\_utils} module in SuperSCRAM. The \texttt{PolygonGeo} class provides a polygonal geometry with great circle arc sides which uses these analytic calculations, and the \texttt{PolygonPixelGeo} class provides the pixelated equivalent.

\section{Galaxy Number Density}
\label{app:nz_candels}

To simulate a sample redshift distribution for a future weak lensing survey, we take all galaxies in the CANDELS \cite{candels_1,candels_2} GOODS-S catalogue with i band magnitude below a selected cutoff (i<24 is used as a default). We then calculate a simple smoothed number density as a function of redshift using a Gaussian smoothing kernel with user specified width $\sigma$ and reflecting boundary conditions at $z=0$:
\begin{align}\label{candels_dndzofz}
\frac{dN}{dz d\Omega}(z)=\sum_{i}{\frac{1}{\sqrt{2\pi\sigma\Omega}}\left(e^{-\frac{(z-z_i)^2}{2\sigma^2}}+e^{-\frac{(z+z_i)^2}{2\sigma^2}}\right)},
\end{align}
where $\Omega$ is the area of the CANDELS survey. Then we can calculate $n(z)$:
\begin{align}\label{candels_nofz}
n(z) = \frac{1}{r(z)^2}\frac{dN}{dz d\Omega}(z)\frac{dz}{dr}(z).
\end{align}
Now, we want to use $n(z)$ to calculate $b(z)$, the bias. Because we are simply demonstrating a possible mitigation strategy and not attempting to model it in detail, we adopt a simple approximation in which we use this $n(z)$ to extract a $M_\text{min}$ for the halo mass function and use it to calculate the bias, neglecting any subtleties from the mass-luminosity relationship. We use the Sheth-Tormen halo mass function \cite{sheth_tormen_hmf,hmf_hu,hmf_reed},
\begin{align}\label{hmf_f}
n(M_\text{min},z)=\int_{M_\text{min}}^\infty{\frac{dn}{dM}dM}=\int_{M_\text{min}}^\infty\frac{\bar{\rho}}{M}\frac{d \ln{\sigma^{-1}}}{dM}f(\sigma)dM,
\end{align}
where $\bar{\rho}$ is the background matter density in units of $h^2 M_\odot \text{Mpc}^{-3}$,
 and $\sigma(M,z)$ is given by 
\begin{align}
\sigma^2(M,z)=\frac{G^2(z)}{2\pi^2G_0}\int_0^\infty k^2 dk P_\text{lin}(k)W^2(k,M),
\end{align}
where $W(k,M)=W(k,R)= 3j_1(kr)/(kr)$
is the Fourier transform of a spherical top hat window function, and $R=\frac{1}{h}[{3M}/(4\pi\bar{\rho})]^{1/3}$, $P_\text{lin}(k)$ is the linear power spectrum and $G(z)$ is the linear growth factor. The Sheth-Tormen mass function itself is
\begin{align}\label{f_shethtormen}
f(\sigma) = A\sqrt{\frac{2a}{\pi}}\left(1+\left(\frac{\sigma^2}{a\delta_c^2}\right)^p\right)e^{-a\frac{\delta_c^2}{\sigma^2}},
\end{align}
where $A=0.3222$, $a=0.707$, and $p=0.3$ are empirically fit parameters, and we take $\delta_c=1.686$ to be the critical overdensity for spherical collapse to avoid considering any cosmology dependence. From here numerically solve $n(M_\text{min},z)=n(z)$ to obtain $M_\text{min}$. Then to calculate the linear bias, we use 
\begin{align}\label{b_shethtormen}
b(\sigma)= 1+\frac{\frac{a\delta_c^2}{\sigma^2}-1}{\delta_c}+\frac{2p}{\delta_c\left(1+\left(\frac{a\delta_c^2}{\sigma^2}\right)^p\right)},
\end{align}
where $\sigma^2=\sigma^2(M_\text{min},z)$. SuperSCRAM could be extended to use improved fitting functions such as in refs.~\cite{jenkins_hmf} or \cite{hmf_tinker}, although the Sheth-Tormen functions are sufficient for the demonstration in this paper.

In SuperSCRAM, eqs.~\eqref{b_shethtormen} and \eqref{candels_nofz} are calculated by the \texttt{NZCandel} class.

\section{Growth Factor}
\label{app:growth_factor}

Simply modifying the calculation of the linear growth factor accounts for most of the correction to the matter power spectrum from dark energy with a variable equation of state. In general relativity, the linear growth factor $G(a)$ evolves according to 
\begin{align}\label{growth_time}
0=G''(a)a^2 H^2(a)+G'\left(\ddot{a}+2a H^2(a)\right)-\frac{3}{2}\frac{\Omega_m H_0^2}{a^3}G,
\end{align}
where the primes denote derivatives with respect to a and the dots are derivatives with respect to cosmic time, and $H(a)$ is the Hubble rate \cite{weinberg_probes}. In the presence of an arbitrary set of perfect fluids with densities $\{\Omega_i(a)\}$ and equations of state $\{w_i(a)\}$, the Hubble rate is
\begin{align}\label{hubble_expansion}
\frac{H(a)^2}{H_0^2}=\sum_i{\Omega_i e^{3\int_a^1(1+w_i(a'))\frac{1}{a'}da'}}
\end{align}
and we can use $D(a)\equiv\frac{G(a)}{a}$ to arrive at a differential equation for $D(a)$:
\begin{align}\label{growth_diff}
D''(a)=-\frac{1}{a}D'(a)\sum_i{\left(\frac{7}{2}-\frac{3}{2}w_i(a)\right)\Omega_i(a)}-\frac{1}{a^2}D(a)\sum_i{\frac{3}{2}\left(1-w_i(a)\right)\Omega_i(a)}
\end{align}
these equations take simple forms for constant w solutions, such as matter, with $w_m=0$, radiation with $w_r=\frac{1}{3}$, and curvature $w_k=-\frac{1}{3}$, and a cosmological constant $w_\Lambda=-1$. For dark energy, there are closed form solutions to the integral $e^{3\int_a^1(1+w(a'))}$ for all dark energy parametrizations of $w(a)$ considered in this paper. SuperSCRAM can be extended to other parametrizations by simply providing functions returning $w(a)$ and $e^{3\int_a^1(1+w(a'))}$. The \texttt{CosmoPie} class can then evaluate $D(a)$ using SciPy's \texttt{odeint} to solve eq.~\eqref{growth_diff}. 

\section{Code Tests}

We have performed a number of tests to verify the various modules of SuperSCRAM perform as expected. Most of the code is covered by unit tests using \texttt{pytest}.

The \texttt{FisherMatrix} class and its associated \texttt{algebra\_utils} module have unit tests for all their functions in \texttt{fisher\_tests.py} and \texttt{algebra\_tests.py}, which run every test with a suite of input matrices, including randomly generated matrices. For \texttt{FisherMatrix}, every test is also run for \texttt{FisherMatrix} objects with every possible internal state, to verify the external behavior of the functions is independent of internal state, as described in subsection~\ref{ssec:fisher_objects}.

The \texttt{PolygonGeo}, \texttt{PolygonPixelGeo}, and \texttt{RectGeo} classes have unit tests in \texttt{polygon\_geo\_tests.py} which verify that their results are both consistent between classes for the same geometries, and that their calculated $a_{lm}$ correctly describe the window function for the input survey geometry. Note that if results for $l>85$ are needed, SuperSCRAM requires the arbitrary precision \texttt{mpmath} package to avoid numerical overflows at double precision.  

The \texttt{power\_response} module's predictions for $\frac{\partial P^{\delta\delta}}{\partial{\bar{\delta}}}$ are compared to the results of \cite{chiang_response} in \texttt{power\_response\_test.py}, and qualitatively agree at the level expected given that they have convolved their power spectrum with a window function and we have not.

The response of a shear shear lensing observable $C^{\gamma\gamma}_l$ to a density perturbation as discussed in  subsection~\ref{ssec:obs_response} and Appendix~\ref{app:su_response} and calculated by \texttt{ShearPower} should have functional dependences of the form $\frac{\partial C^{\gamma\gamma}_l}{\partial\bar{\delta}}(z_s)\propto\frac{\Delta z_s}{z_i}C^{\gamma\gamma}_l$, where $\Delta z_s$ is the width of a resolution $z$ slice integrated over in \eqref{chain_obs} and $z_i$ is the average $z$ of the tomographic bin. The functional dependences of $\frac{\partial C^{\gamma\gamma}_l}{\partial\bar{\delta}}(z_s)$ are checked by \texttt{power\_derivative\_tests.py}.

Our Python implementation of Halofit used by the \texttt{MatterPowerSpectrum} class described in subsection~\ref{ssec:matter_power} is very similar to the Fortran implementation of the Takahashi Halofit prescription available in CAMB \cite{camb,takahashi_halofit}, with some modifications to facilitate our prescriptions for arbitrary $w(z)$, and computing $\frac{\partial P^{\delta\delta}}{\partial\bar{\delta}}$. The agreement with the CAMB output is tested in \texttt{power\_comparison\_tests.py}. Our implementation agrees with CAMB's implementation of the Takahashi Halofit prescription to within $0.2\%$ for $k\lesssim10 $ Mpc$^{-1}$. Most of the residual difference is because we blend the transition between the linear and nonlinear power spectrum to avoid sharp spikes in $\frac{\partial P^{\delta\delta}}{\partial\bar{\delta}}$ in eq.~\eqref{chain_obs}, while CAMB's implementation uses a sharp transition at $k=0.005 h$ Mpc$^{-1}$.

Our prescription for an arbitrary $w(a)$ prescription implemented in the \texttt{WMatcher} class, discussed in Appendix \ref{app:modify_matter} is tested in \texttt{w\_matcher\_tests.py}. We test that our results for our \texttt{w0wa} model with $w(a)=w_0+(1-a)w_a$ match the results of \cite{casarini_halofit_math}, and that our 36 bin \texttt{jdem} model gives results consistent with the \texttt{w0wa} prescription for similar $w(a)$. We also test that both \texttt{w0wa} and \texttt{jdem} models correctly recover the known $w_0$ for constant $w(a)$ models. 

The variance in a window with window function $W({\bf k})$ is given $\sigma^2=\int \frac{d^3 {\bf k}}{(2\pi)^3} P({\bf k}) W({\bf k})$ and can be written in our formalism $\sigma^2=\frac{\partial\bar{\delta}}{\partial\delta^\alpha}C^{\alpha\beta}\frac{\partial\bar{\delta}}{\partial\delta^\beta}$. The script \texttt{variance\_demo.py} compares results from SuperSCRAM using \texttt{PolygonGeo} to a code which integrates the linear matter power spectrum directly for a rectangular window function. For an approximately cubic $500 Mpc$ window, SuperSCRAM converges to within about $2\%$ of the $\sigma^2$ predicted by SuperSCRAM which directly integrates the matter power spectrum. The \texttt{variance\_demo.py} module also enables convergence testing our basis decomposition. In general the decomposition converges faster when the volume of the geometry is increased, or when $z_\text{max}$ is decreased. 

The script \texttt{super\_survey\_tests.py} does a full run of the sequence necessary to run SuperSCRAM, including various consistency checks on the calculated eigenvalues. 

In addition to the testing modules described here, we have verified by conducting multiple runs that the results converged to ${\cal O}(0.01\%)$ with respect to all the various parameters, except for the number of basis elements. 

\section{Modifying the Matter Power Spectrum}
\label{app:modify_matter}

Motivated by the procedure in ref.~\cite{casarini_halofit_math}, for a given $w(z)$, the \texttt{w\_matcher} module calculates an effective equation of state for dark energy $\mathcal{W}(z)$ which represents, at a given $z$, the constant $w_{\text{eff}}$ which reproduces the same comoving distance to last scattering,
\begin{align}\label{cas_cond1}
\int_z^{z_{\text{lss}}}\frac{d z'}{E(z',w=\mathcal{W}(z))}=\int_z^{z_{\text{lss}}}\frac{d z'}{E(z',w=w(z))}
\end{align}
where $E(z)=H(z)/H_0$ and $z_{\text{lss}}$ is the redshift of last scattering. Currently, the \texttt{WMatcher} class precomputes the left-hand side of eq.~\eqref{cas_cond1} on a grid of possible $w$ and $z$ values and interpolates to match the right-hand side. Then, the amplitude of the $z=0$ linear matter power spectrum must be rescaled using $P_{\text{lin}}(k,z=0)\rightarrow\mathcal{G}^2(z)P_{\text{lin}}(k,z=0)$ where $\mathcal{G}^2(z)$ is obtained from 
\begin{align}\label{cas_cond2}
\mathcal{G}^2(z)\left(\frac{G(z,w=\mathcal{W}(z))}{G(0,w=\mathcal{W}(z))}\right)^2=\left(\frac{G(z,w=w(z))}{G(0,w=w(z))}\right)^2.
\end{align}
Note that this condition is equivalent to eq.~(2.3) in ref.~\cite{casarini_halofit_math}, but this form is clearer in parametrizations where $\sigma_8$ is not a fixed parameter. The \texttt{w\_matcher} module matches this condition by precomputing a grid of possible $G(z,w)$ values and interpolating for a given $\mathcal{W}(z)$. Once $\mathcal{W}(z)$ is evaluated, the model is treated as having that constant $w$ value in all respects for calculations involving that $z$ value; for example, we must obtain a new linear power spectrum, because the small $k$ transfer function depends on $w$.

\section{Separate Universe Response}
\label{app:su_response}
The response of a the matter power spectrum to a long-wavelength density fluctuation can be approximated by treating the overdense region as a 'separate universe', which can be used to find the response of an observable to density fluctuations. The response of the linear matter power spectrum to a fluctuation $\bar{\delta}(t)$ is \cite{chiang_response,baldauf_response,baldauf_linear,lihu_ssc_sim}

\begin{equation}\label{linear_response}
\frac{d\ln P^{\delta\delta}_{\text{lin}}(k,a)}{d\bar{\delta}(t)}=\frac{68}{21}-\frac{1}{3}\frac{d\ln k^3 P^{\delta\delta}_{\text{lin}}(k,a)}{d\ln k}.
\end{equation}
The one-loop correction simply picks up an additional factor of $\frac{26}{21}$,
\begin{equation}\label{1-loop_response}
\frac{d\ln P^{\delta\delta}_{\text{1-loop}}(k,a)}{d\bar{\delta}(t)}=\frac{68}{21}-\frac{1}{3}\frac{d\ln k^3 P^{\delta\delta}_{\text{1-loop}}(k,a)}{d\ln k}+\frac{26}{21}\frac{P^{\delta\delta}_{\text{1-loop}}(k,a)-P^{\delta\delta}_{\text{lin}}(k,a)}{P^{\delta\delta}_{\text{1-loop}}(k,a)},
\end{equation}
and for the Halofit power spectrum, we follow the prescription in ref.~\cite{chiang_response} by absorbing the factor of $\widetilde{G}(\tilde{a})=\left(1+\frac{13}{21}\bar{\delta}(t)\right)G(a)$ into $\tilde{\sigma}_8=\left(1+\frac{13}{21}\bar{\delta}(t_0)\right)\sigma_8$, so that 
\begin{equation}\label{halofit_response}
\frac{d\ln P^{\delta\delta}_{\text{Halofit}}(k,a)}{d\bar{\delta}(t)}=\frac{13}{21}\frac{d\ln P^{\delta\delta}_{\text{Halofit}}(k,a)}{d\ln\sigma_8}+2-\frac{1}{3}\frac{d\ln k^3 P^{\delta\delta}_{\text{Halofit}}(k,a)}{d\ln k}.
\end{equation}

Because at present we only consider $w(z)$ models as perturbations around $w=$constant models, we can ignore any correction to these terms due to a variable $w(z)$. The response of the shear-shear lensing power spectrum can be approximated by plugging the response of the separate universe matter power spectrum into eq.~\eqref{limber_cab}:

\begin{equation}\label{lensing_response}
\frac{\partial C^{AB}(\ell)}{\partial\bar{\delta}}\cong\int_{0}^{r_{max}}{dr\, \frac{q^A(r)q^B(r)}{r^2}\frac{\partial P^{\delta\delta}}{\partial\bar{\delta}}\left(k=\frac{\ell+1/2}{r}\right)}.
\end{equation}

The response of the observable to fluctuations in our basis, $\frac{\partial C^{AB}(\ell)}{\partial\delta^\alpha}$ can then be calculated using the chain rule as described in subsection~\ref{ssec:obs_response}, obtaining:
\begin{equation}\label{lensing_basis_response}
\frac{\partial C^{AB}(\ell)}{\partial\delta^\alpha}\cong\int_{0}^{r_{max}}{dr\, \frac{q^A(r)q^B(r)}{r^2}\frac{\partial P^{\delta\delta}}{\partial\bar{\delta}}\left(k=\frac{\ell+1/2}{r}\right)}\frac{\partial\bar{\delta}}{\partial\delta^\alpha}(r).
\end{equation}

\end{document}